\documentclass{aa}

\usepackage{graphicx}
\usepackage{rotating}
\usepackage{txfonts}
\usepackage{threeparttable}
\usepackage{float}
\usepackage{xcolor}
\usepackage{pdflscape}
\usepackage{tablefootnote}
\usepackage{textcomp, gensymb}
\usepackage{gensymb}
\usepackage[colorlinks=true, linkcolor=blue, citecolor=blue, filecolor=blue, urlcolor=blue]{hyperref}
\usepackage{dblfloatfix}
\usepackage{placeins}
\usepackage{longtable}
\usepackage{caption}

\begin{document}

\title{Mass determination of the ultra-short-period planet LHS~3844~b}
\subtitle{First $K$-band radial velocity measurements with CRIRES$^+$\thanks{Based on guaranteed time observations (GTO) 
collected at the European Southern Observatory (ESO) under ESO programs 108.22CH.001, 109.23FA.001, 110.2447.001, and 
111.24SC.001 by the CRIRES$^+$ consortium.}\fnmsep\thanks{Based on observations collected at ESO under ESO programs 0102.C-0496(B) and 0103.C-0849(A).}}

\author{
     E.~Nagel\inst{\ref{inst:iag}}
\and J.~Köhler\inst{\ref{inst:tls}}
\and M.~Zechmeister\inst{\ref{inst:iag}}
\and A.~D.~Rains\inst{\ref{inst:upp},\ref{inst:chi}}
\and U.~Seemann\inst{\ref{inst:eso_g}}
\and A.~Hatzes\inst{\ref{inst:tls}}
\and A.~Reiners\inst{\ref{inst:iag}}
\and N.~Piskunov\inst{\ref{inst:upp}}
\and L.~Boldt-Christmas\inst{\ref{inst:upp}}
\and P.~Bristow\inst{\ref{inst:eso_g}}
\and P.~Chaturvedi\inst{\ref{inst:tls}}
\and D.~Cont\inst{\ref{inst:lmu}}
\and S.~Czesla\inst{\ref{inst:tls}}
\and R.~J.~Dorn\inst{\ref{inst:eso_g}}
\and E.~Guenther\inst{\ref{inst:tls}}
\and Y.~Jung\inst{\ref{inst:eso_g}}
\and O.~Kochukhov\inst{\ref{inst:upp}}
\and F.~Lesjak\inst{\ref{inst:aip}}
\and F.~Lucertini\inst{\ref{inst:eso_c}}
\and T.~Marquart\inst{\ref{inst:upp}}
\and L.~Nortmann\inst{\ref{inst:iag}}
\and M.~Rengel\inst{\ref{inst:mps}}
\and F.~Rodler\inst{\ref{inst:eso_c}}
\and J.~V.~Smoker\inst{\ref{inst:eso_c}, \ref{inst:atc}}
}

\institute{
\label{inst:iag} Universit\"at G\"ottingen, Institut f\"ur Astrophysik und Geophysik, Friedrich-Hund-Platz 1, 37077 G\"ottingen, Germany\\
\email{evangelos.nagel@uni-goettingen.de}
\and \label{inst:tls} Th\"uringer Landessternwarte Tautenburg, Sternwarte 5, 07778 Tautenburg, Germany
\and \label{inst:upp} Department of Physics and Astronomy, Uppsala University, Box 516, 75120 Uppsala, Sweden
\and \label{inst:chi} Instituto de Astrofísica, Pontificia Universidad Católica de Chile, Av. Vicuña Mackenna 4860, 782-0436 Macul, Santiago, Chile
\and \label{inst:eso_g} European Southern Observatory, Karl-Schwarzschild-Str. 2, 85748 Garching bei München, Germany
\and \label{inst:lmu} Universitäts-Sternwarte, Ludwig-Maximilians-Universität München, Scheinerstrasse 1, 81679 München, Germany
\and \label{inst:aip} Leibniz Institute for Astrophysics Potsdam (AIP), An der Sternwarte 16, 14482 Potsdam, Germany
\and \label{inst:eso_c} European Southern Observatory, Alonso de Cordova 3107, Vitacura, Casilla 19001, Santiago, Chile
\and \label{inst:mps} Max-Planck-Institut für Sonnensystemforschung, Justus-von-Liebig-Weg 3, 37077 Göttingen, Germany
\and \label{inst:atc} UK Astronomy Technology Centre, Royal Observatory, Blackford Hill, Edinburgh EH9 3HJ, UK
}

\date{Received 27 October 2025 / Accepted 13 March 2026}

\abstract
{We present the first planet mass measurement obtained with CRIRES$^+$ radial velocity (RV) observations using the $K$-band gas cell. Our target, LHS~3844~b (TOI-136), is a transiting super-Earth with radius $R_{\rm b}=1.286^{+0.043}_{-0.044}\,R_\oplus$ and an orbital period of $P_{\rm b} = 0.462929709^{+0.000000044}_{-0.000000042}$\,d, placing it in the class of ultra-short-period (USP) planets. The host star LHS~3844 is an old ($7.8\pm1.6$\,Gyr), slowly rotating ($P_{\rm rot} = 130.0^{+16.9}_{-13.4}$\,d) M5.0 dwarf with $M_\star = 0.151\pm0.014\,M_\odot$ at a distance of $15$\,pc ($V=15.2\,$mag, $K=9.2\,$mag). Combining our CRIRES$^+$ RVs with archival ESPRESSO spectra, and confirming the signal in each dataset independently, we detected periodic RV variations with a semi-amplitude $K_{\rm b}=6.95^{+0.55}_{-0.60}$\,m\,s$^{-1}$, implying a planetary mass of $m_{\rm b} = 2.37\pm0.25\,M_\oplus$ and a bulk density of $\rho_{\rm b} = 6.15^{+0.60}_{-0.61}$\,g\,cm$^{-3}$, consistent with a predominantly rocky composition. We further found excess RV variability that may be attributed to stellar jitter or to an additional planetary signal, for which we identified a tentative super-Earth candidate with a period of $\approx6.88\,$d.
Owing to its proximity to its M-dwarf host, LHS~3844~b experiences intense irradiation and is unlikely to retain a substantial H/He envelope. 
Interior modeling places an upper limit on the iron-core mass fraction, which is consistent with an Earth-like rocky composition.
With an emission spectroscopy metric (ESM) of 28, LHS~3844~b is a prime JWST target for atmospheric and surface characterization and the most promising surface-characterization target known. Phase-curve spectroscopy may reveal its surface mineralogy and enable the first robust detection of exoplanet surface spectral features. Our results demonstrate that near-infrared RVs obtained with CRIRES$^+$ enable robust mass measurements of super-Earths orbiting late M dwarfs.} 


\maketitle
\nolinenumbers

\section{Introduction}

The advent of exoplanet transit missions such as CoRoT, \textit{Kepler}, and more recently TESS has revealed that planets with sizes and orbital periods unlike those found in the Solar System are quite common. Among these, the so-called ultra-short-period (USP) planets, defined as having orbital periods shorter than one day, constitute a distinct population. A prominent example is LHS~3844~b \citep{Vanderspek2019}, the first USP planet detected by TESS in Sector 1. LHS~3844~b is a tidally locked \citep{Lyu2024} super-Earth with a radius of $1.3\,R_\oplus$ that orbits its M5.0-type host star every $11.11$ hours. 

The atmospheric properties of LHS~3844~b have been probed with thermal phase-curve observations from \textit{Spitzer} at 4-5\,$\mu$m \citep{Kreidberg2019}. These data yield a dayside brightness temperature of $1040 \pm 40$\,K and a nightside broadly consistent with $0$\,K, implying negligible heat redistribution. Such an extreme hemispheric temperature contrast disfavors the presence of a substantial atmosphere, effectively ruling out a thick atmosphere with surface pressures exceeding \mbox{$\sim10$\,bar}. Independent optical transmission spectroscopy provides further support for this interpretation \citep{Diamond-Lowe2020}. Similar results for other terrestrial USPs from \textit{Spitzer} and \textit{JWST} phase-curve emission spectroscopy bolster the ``bare rock'' interpretation \citep{Crossfield2022, Greene2023, Zieba2023, WeinerMansfield2024, Xue2024, Zhang2024, Allen2025, Ducrot2025, Luque2025, Wachiraphan2025}. 

These non-detections support the interpretation that USP planets may represent stripped rocky cores. Owing to their close proximity to their host stars, USP planets are exposed to intense stellar irradiation. Models predict that those with sufficiently low masses can lose their primordial H/He envelopes through photo-evaporation driven by intense extreme ultraviolet (XUV) irradiation of their host star \citep{Lundkvist2016, Owen2013, Owen2017}, leaving behind a bare solid core. Such objects provide a rare opportunity to directly constrain planetary composition. With an emission spectroscopy metric (ESM) of 28 \citep{Kempton2018}, LHS~3844~b is an exceptionally favorable target for emission studies, placing it among the strongest emission sources known for exoplanets with solid rock daysides, and making it one of the most compelling targets for probing surface properties and geologic history. Accordingly, three secondary eclipses with the Mid-Infrared Instrument low-resolution spectroscopy mode (MIRI/LRS) were obtained in Cycle 1 (PI: Kreidberg), followed by a Near-Infrared Spectrograph (NIRSpec) phase curve in Cycle 2 (PI: Zieba) and nine additional MIRI/LRS visits in Cycle 4 (PI: Paragas), totaling $50.3$\,h of \textit{JWST} time. Precise measurements of mass and radius are therefore essential to interpret these data and, more generally, to infer interior structure, bulk composition, and surface properties, yielding key insights into the formation and evolution of terrestrial exoplanets.

This paper is dedicated to the mass measurement of LHS~3844~b, representing the first such determination based on radial velocity (RV) observations obtained with the upgraded CRyogenic InfraRed Echelle Spectrograph \citep[CRIRES$^+$;][]{Dorn2023} in conjunction with the Echelle SPectrograph for Rocky Exoplanets and Stable Spectroscopic Observations \citep[ESPRESSO;][]{Pepe2021}. Our results provide the bulk-density anchor needed to study the system's composition and surface geology.

\section{Observations}
\label{sect:observations}

\subsection{TESS photometry}
\label{sect:tess_photometry}

Launched in April 2018, the TESS spacecraft commenced its regular science operations in July 2018 \citep{Ricker2015}.
The core mission aims to conduct a near all-sky survey with the goal of discovering exoplanets with masses smaller than Neptune as they transit stars in close proximity to Earth. TESS is equipped with four optical cameras, which together provide a combined field of view spanning $24\degree \times 96\degree$. 

LHS~3844 was monitored by TESS in Sector~1 during its primary mission with a 2-minute cadence. During the first two extended missions, observations were made in Sectors 27, 28, 67, and 68, employing both 20-second and 2-minute cadences. The calibrated light curves, processed by the Science Processing Operations Center \citep[SPOC;][]{Jenkins2016}, were retrieved from the Mikulski Archive for Space Telescopes\footnote{\url{https://mast.stsci.edu}} (MAST). The photometric data encompassed simple aperture photometry (SAP) and pre-search data conditioning SAP (PDCSAP) flux measurements. The PDCSAP data were specifically processed to mitigate systematic trends and artifacts while preserving planetary transit signals.

\begin{figure}
\begin{center}
\includegraphics[width=0.49\textwidth]{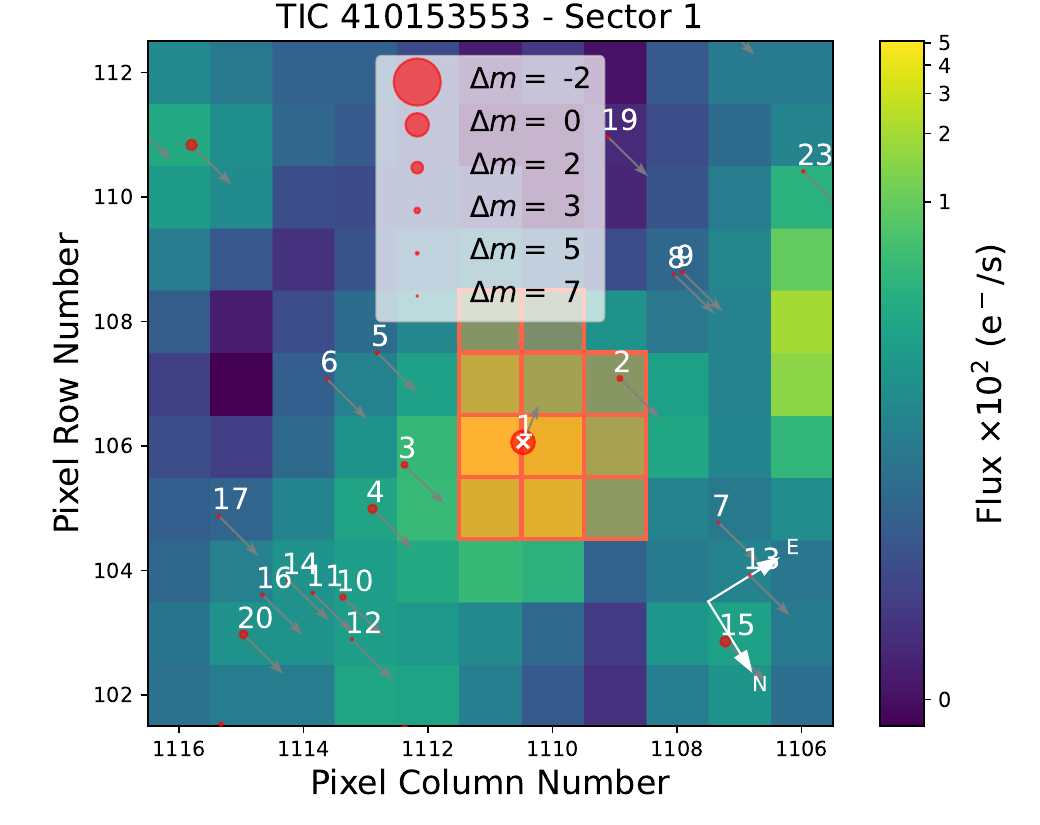}
\caption{Target pixel file image for LHS~3844 from TESS Sector~1, with electron counts represented through color coding. The aperture mask applied by the SPOC pipeline for SAP extraction is highlighted by red squares. Nearby objects cataloged in Gaia DR3 are denoted by red circles, the sizes of which scale according to the magnitude contrast with LHS~3844. The central position of LHS~3844 is marked by a white cross, and the arrows denote proper motion directions.}
\label{figure:TPF}
\end{center}
\end{figure}

By examining the target pixel files (TPF) from the five Sectors, we inspected the apertures and searched for potential sources of contamination, given that a single TESS pixel covers a relatively large area of approximately 21 square arcseconds. In Fig.~\ref{figure:TPF} we show the TPF image centering LHS~3844 from Sector 1 generated with the \texttt{tpfplotter}\footnote{\url{https://github.com/jlillo/tpfplotter}} tool \citep{Aller2020}. Over-plotted are sources from the Gaia DR3 catalog \citep{GaiaDR32023} up to a Gaia $G$-band magnitude contrast of $\Delta m = 7\,$mag. Among the five Sectors analyzed, we identified that Sector 28 was devoid of any contamination. However, Sectors 1, 27, and 68 were found to be contaminated by source 2 ($\Delta m = 4.4\,$mag, projected separation $r=39\,$arcsec), while Sector 67 exhibited contamination from source 3 ($\Delta m = 3.9\,$mag, $r=41\,$arcsec). To evaluate potential contamination of the light curves by these sources, we analyzed the crowding metric generated by the SPOC pipeline. This metric approximates the fraction of flux within the optimal aperture that originates from the target star. For LHS~3844, the SPOC metric is in the range of $97.8 \text{-} 99.5\,\%$ for all Sectors, which indicates that this fraction of the flux in the aperture belongs to LHS~3844 and $2.2 \text{-} 0.5\,\%$ originates from other objects. The SPOC pipeline corrects the PDCSAP light curves for this excess flux, which may lead to shallower transit depths and, consequently, a systematic underestimation of planetary radii \citep{Stumpe2012}. The observation log for the TESS data is provided in Table~\ref{table:photometry} and the full PDCSAP light curves for all Sectors are shown in Fig.~\ref{figure:juliet_fit_TESS_GPDetrending}.

\begin{table*}
\begin{center}
\caption{Observing log for TESS data on LHS~3844.}
\label{table:photometry}
\begin{tabular}{ccccccc} 
\hline\hline 
\noalign{\smallskip}
Start date & End date & Cycle & Sector & Camera & Cadence & \# of transits \\
\noalign{\smallskip}
\hline
\noalign{\smallskip}
2018-07-25 & 2018-08-22 & 1 &  1 & 3 & 2\,min & 57 \\
2020-07-05 & 2020-07-30 & 3 & 27 & 3 & 20\,s & 49 \\
2020-07-31 & 2020-08-25 & 3 & 28 & 3 & 20\,s & 46 \\
2023-07-01 & 2023-07-28 & 5 & 67 & 2 & 20\,s & 39 \\
2023-07-29 & 2023-08-25 & 5 & 68 & 2 & 20\,s & 44 \\
\noalign{\smallskip}
\hline
\end{tabular}
\end{center}
\end{table*}

\begin{table*}
\begin{center}
\caption{Observing log for the CRIRES$^+$ and ESPRESSO RV measurements of LHS~3844.}
\label{table:rv}
\begin{tabular}{ccccccc} 
\hline\hline 
\noalign{\smallskip}
Instrument & Baseline & Measurements & $\sigma_{\rm eff}$ [m\,s$^{-1}$] & WRMS [m\,s$^{-1}$] & Residual WRMS [m\,s$^{-1}$]\\ 
\noalign{\smallskip}
\hline
\noalign{\smallskip}
ESPRESSO preUp & 2018-11-02 -- 2018-12-03 & 13 & 0.53 & 5.44 & 1.52 \\ 
ESPRESSO postUp & 2019-07-11 -- 2019-09-06 & 10 & 0.46 & 5.96 & 1.99 \\ 
CRIRES$^+$ & 2021-10-09 -- 2023-09-04 & 60 & 4.35 & 5.81 & 3.05 \\ 
\noalign{\smallskip}
\hline
\end{tabular}
\tablefoot{
$\sigma_{\rm eff}$ denotes the effective per-measurement uncertainty (Sect.~\ref{sect:rv_data_quality}), WRMS is the error-weighted root-mean-square calculated relative to the weighted mean, and the residual WRMS is the error-weighted RMS of the $\mathrm{O - C}$ residuals (Sect.~\ref{section:joint}).}
\end{center}
\end{table*}

\subsection{MEarth photometry}

Photometric variability induced by surface inhomogeneities, such as cool spots and bright plages, allows the stellar rotation period to be inferred through the analysis of periodic brightness modulation. To investigate this variability, we retrieved archival light curves from the MEarth-South\footnote{\url{https://lweb.cfa.harvard.edu/MEarth/Welcome.html}} survey \citep{Berta2012}. MEarth-South, located at the Cerro Tololo Inter-American Observatory (CTIO) in Chile, comprises eight robotically operated 40\,cm telescopes. Each telescope is equipped with a $2048 \times 2048$ CCD, offering a pixel scale of $0.84^{\prime\prime}$ resulting in a field of view of $28.7^{\prime}$, and utilizes a custom 715\,nm long-pass filter. The MEarth project was specifically designed to monitor mid-to-late M dwarfs, targeting a sample of over 1\,000 objects \citep{Irwin2015}. Its primary goal was to detect transiting exoplanets with radii as small as 2\,$R_\oplus$ and orbital periods of up to 20 days, encompassing the habitable zones of these low-mass stars \citep{Nutzman2008}. The survey concluded its operations in February 2022.

LHS~3844 was monitored by MEarth-South between January 2016 and December 2018, yielding a total of 3525 individual photometric measurements. To mitigate the impact of outliers caused by poor weather conditions or instrumental anomalies, we applied a 2.5$\sigma$ clipping procedure. This filtering step removed 29 datapoints, resulting in a cleaned dataset of 3496 measurements. The exposure times of 47 seconds yield a median photometric uncertainty of $\tilde{\sigma}_{\rm MEarth} = 2.7\,$mmag. These observations span 256 epochs, providing a robust time baseline for variability and rotational period analysis.

\subsection{High-resolution spectroscopy}

\subsubsection{CRIRES$^+$}

\begin{figure*}  
\begin{center}
\includegraphics[width=1.0\textwidth]{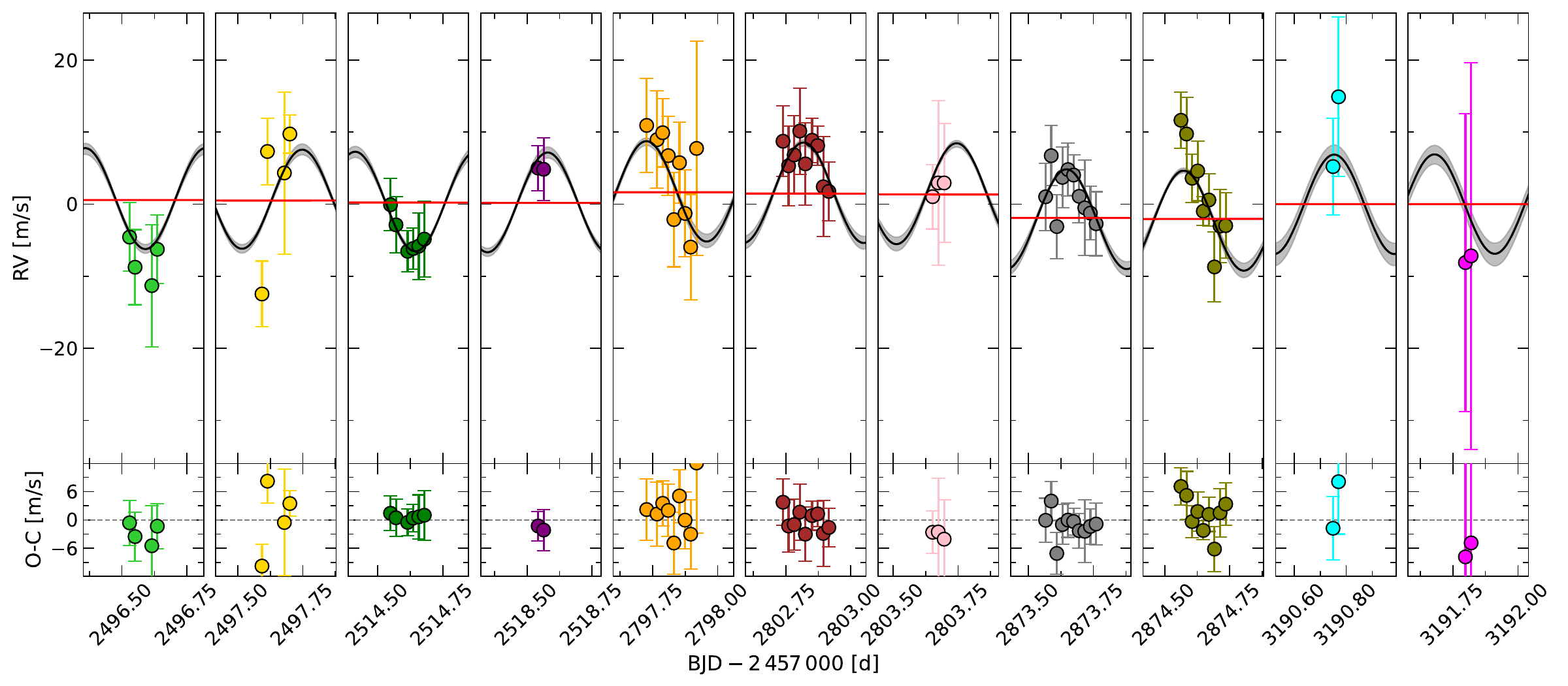}
\caption{\label{figure:crires_rv}
\textit{Top panels:} Radial velocity measurements of LHS~3844 obtained with CRIRES$^+$. The solid black line indicates the best-fit Keplerian model for a single planet on a circular orbit (Sect.~\ref{section:joint}), with the gray shaded region marking the 68\,\% confidence interval. The red solid line shows the GP component. Each panel covers exactly one orbital period of LHS~3844~b and displays data from the 11 observing runs, color-coded accordingly. The large uncertainties in the rightmost panel are due to poor weather conditions.
\textit{Bottom panels}: ${\rm O} - {\rm C}$ residuals.}
\end{center}
\end{figure*}

Spectroscopic observations were carried out using the CRIRES$^+$ echelle spectrograph \citep{Dorn2023}, installed at the Nasmyth B focus of the UT3 8\,m telescope of the Very Large Telescope (VLT) at ESO's Paranal Observatory in Chile. CRIRES$^+$ covers the $Y$ to $M$ bands (0.95-5.3\,$\mu$m) and delivers a spectral resolving power of $\mathcal{R} \simeq 100\,000$ or $50\,000$, depending on the use of the $0.2^{\prime\prime}$ or $0.4^{\prime\prime}$ slit, respectively. The instrument is equipped with three $2048 \times 2048$ Hawaii-2RG detectors and a cross-disperser, enabling up to ten times greater instantaneous wavelength coverage compared to its predecessor. With its enhanced stability and coverage, CRIRES$^+$ is well suited for one of its primary upgrade science cases, the study of exoplanets around M dwarfs\footnote{\url{https://www.eso.org/sci/facilities/develop/instruments/crires_up.html}}.

We obtained 60 RV measurements of LHS~3844 with CRIRES$^+$ over a baseline of 695 days. Given the planet's orbital period of $\sim 11.1$\,h, a substantial portion of the orbit can be sampled within a single night. Observations were therefore conducted over 11 nights between October 2021 and September 2023 as part of the CRIRES$^+$ Guaranteed Time Observations (GTO) survey under ESO program IDs 108.22CH.001, 109.23FA.001, 110.2447.001, and 111.24SC.001 (PI: Nagel). Notably, LHS~3844~b was the first exoplanet targeted by the CRIRES$^+$ consortium for measuring the mass with the RV method. The observation log of the CRIRES$^+$ data is given in Table~\ref{table:rv}.

We employed a slit width of $0.2^{\prime\prime}$ to maximize the spectral resolving power. However, as reported by \citet{Dorn2023}, the resolving power during the commissioning runs was significantly lower ($\mathcal{R}\gtrsim 80\,000$) than the nominal value of $\mathcal{R} \approx 100\,000$. This reduced resolution affected the first 16 RV measurements in our dataset. Following an optimization of the camera focus in March 2022, the resolving power of CRIRES$^+$ improved to at least $\mathcal{R} \approx 92\,000$ in the wavelength setting used for our observations.

To effectively subtract the sky background and correct for detector artifacts, we employed a standard nodding strategy, alternating the target between two positions (A and B) along the slit. Observations were carried out using the K2192 wavelength setting, which spans 1946-2501\,nm with spectral gaps between echelle orders. Each nodding position was observed with a detector integration time (DIT) of 900\,s. Additionally, we utilized the Multi-Application Curvature Adaptive Optics system (MACAO) as well as the metrology system.

The absolute wavelength calibration reference for CRIRES$^+$ is provided by Uranium-Neon (UNe) lamp exposures acquired during morning calibrations. Since these UNe exposures cannot be obtained simultaneously with science observations, the wavelength solution produced by the data reduction pipeline may exhibit slight shifts in dispersion relative to the actual science data. These shifts arise from residual metrology alignment errors of approximately $\pm0.1$ pixels and drifts in the echelle grating. The accuracy of nonsimultaneous absolute wavelength calibration in the $K$ band is limited to approximately $500$\,m\,s$^{-1}$, primarily due to the sparse distribution of UNe lines. To improve relative wavelength calibration, we also employed the Fabry-Perot etalon (FPET), enabling nonsimultaneous wavelength calibration in the $K$ band with an accuracy better than $30$\,m\,s$^{-1}$ \citep{Dorn2023}. For high-precision wavelength calibration, we employed the short gas cell (SGC), an absorption cell filled with a mixture of gases that imprints a dense pattern of absorption lines onto the spectra, providing a stable wavelength reference \citep{Seemann2014, Dorn2023}.

We extracted the spectra from the 2D images using the Data Reduction Software (DRS) {\tt CR2RES}\footnote{\url{https://www.eso.org/sci/software/pipelines/cr2res/cr2res-pipe-recipes.html}} (version~1.3.0). The reduction process included flat-fielding, order tracing, slit tilt correction, blaze function extraction, and wavelength calibration using UNe and FPET exposures. By applying the {\tt cr2res\_obs\_nodding} recipe to the science data, we obtained a coadded spectrum from each AB nodding pair.

We computed RVs using our custom pipeline Velocity and IP EstimatoR \citep[\texttt{viper};][]{Koehler2025, Zechmeister2021}, developed by the CRIRES$^{+}$ consortium and publicly available on GitHub\footnote{\url{https://mzechmeister.github.io/viper_RV_pipeline}}. The pipeline constructs a full model of the stellar spectrum as observed through the gas cell, following the technique pioneered by \citet{Valenti1995} and \citet{Butler1996} for the iodine gas cell and later adapted to the $K$ band by \citet{Bean2010} using the former CRIRES spectrograph with the ammonia gas cell. A detailed description of the {\tt viper} pipeline, including its functionality and capabilities, is provided in \citet{Koehler2025}. Therefore, we only summarize the key aspects of the approach.

The model comprises the product of a stellar template and a gas-cell transmission spectrum, both ideally measured at high spectral resolving power and high signal-to-noise (S/N). For the gas-cell spectrum, we use a high-resolution transmission spectrum obtained with a Fourier Transform Spectrograph (FTS) at a resolving power of $\mathcal{R} \approx 1\,000\,000$. This composite model is then convolved with the instrumental profile (IP). Using a least-squares fitting algorithm, {\tt viper} simultaneously optimizes parameters including the wavelength solution, continuum normalization, IP, and stellar Doppler shift. This approach enables effective correction for instrumental instabilities, ensuring the model best matches the observed spectrum. 

\begin{figure*}
\begin{center}
\includegraphics[width=1.0\textwidth]{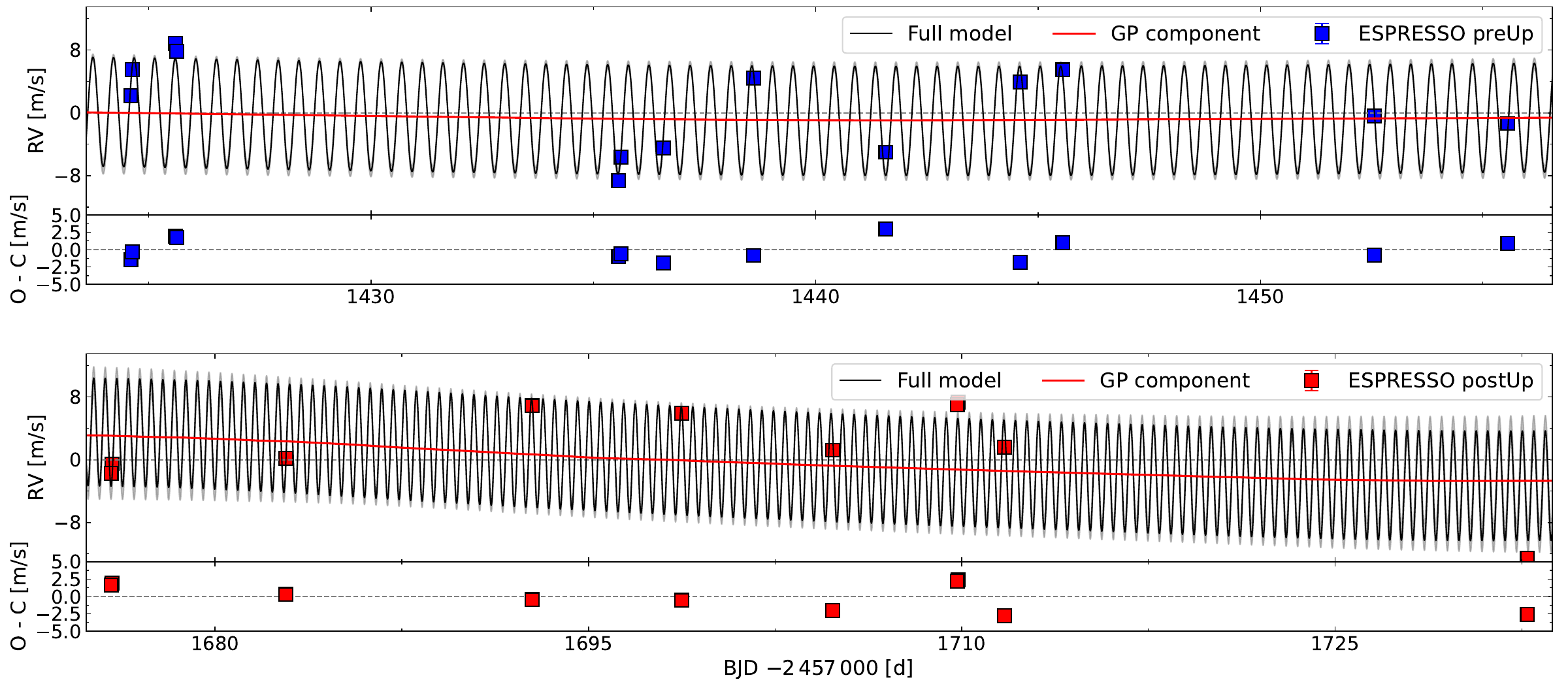}
\caption{\label{figure:espresso_rv}
ESPRESSO RV time series of LHS~3844.
\textit{Upper panels:} 
Radial velocity measurements shown before (blue squares, \textit{top row}) and after (red squares, \textit{bottom row}) the instrumental intervention. The solid black line represents the best-fit Keplerian model for a single planet on a circular orbit (Sect.~\ref{section:joint}), with the gray shaded region marking the 68\,\% confidence interval. The red solid line shows the GP component. Error bars are smaller than the plotted symbols.
\textit{Bottom panels}: ${\rm O} - {\rm C}$ residuals corresponding to the top panels.}
\end{center}
\end{figure*}

To address the pervasive issue of telluric contamination in the near-infrared, {\tt viper} extends this modeling framework by incorporating an additional atmospheric component. Precomputed synthetic telluric spectra for each atmospheric constituent in the $K$ band (CO$_2$, CH$_4$, H$_2$O, N$_2$O, and CO), generated using the Line-By-Line-Radiative-Transfer-Model code \citep[LBLRTM;][]{Clough1995, Clough2005}, are scaled to account for variations in line depth. An additional free parameter is included to correct for the Doppler shift of telluric features, which arises from atmospheric winds and typically corresponds to velocity offsets on the order of 10\,m\,s$^{-1}$ \citep{Figueira2010}. 

In addition, {\tt viper} provides functionality for generating telluric-free stellar templates. In this mode, telluric features are first removed from all observations acquired without the gas cell. These corrected spectra are then coadded using a weighted mean to produce a high S/N stellar template with reduced residual artifacts from imperfect telluric correction. For this work, we constructed a stellar template for LHS~3844 by coadding 12 CRIRES$^+$ spectra taken without the gas cell, each obtained immediately before or after the observations with the gas cell. 
RV shifts are then measured from the gas-cell observations using this template as the reference. For the final RV determination, we used spectral orders in the wavelength range 2130-2470\,nm, which offer the best combination of stellar, gas cell, and telluric lines. The remaining orders were excluded due to strong telluric contamination or a lack of gas-cell lines.
Because key model parameters, such as the IP, wavelength solution, and blaze function, vary with wavelength, RVs are computed separately for each suborder, each covering a wavelength range of approximately 13\,nm. The final RV for each observation is calculated as a weighted mean of the individual order-wise RVs.
Performance tests with {\tt viper} showed that modeling the IP with a Gaussian and fitting the wavelength solution with a second-degree polynomial yielded the best results for our CRIRES$^+$ data. We present the CRIRES$^+$ RV measurements in Fig.~\ref{figure:crires_rv} and list them, together with their formal uncertainties, in Table~\ref{table:crires_rvs}.

\subsubsection{ESPRESSO}

ESPRESSO is a fiber-fed, cross-dispersed, high-resolution echelle spectrograph installed on the VLT array at the ESO Paranal Observatory in Chile \citep{Pepe2021}. ESPRESSO covers a wavelength range of 380-788\,nm and offers a spectral resolving power of $\mathcal{R} \approx 140\,000$ in High-Resolution (HR) mode. To ensure high stability, the instrument is housed within a vacuum vessel stabilized in both temperature and pressure, itself located in a thermally controlled room. This configuration enables long-term RV precision at the level of 10\,cm\,s$^{-1}$. 

A total of 25 individual spectra were collected between November 2018 and September 2019 under ESO program IDs 0102.C-0496 and 0103.C-0849 (PI: Astudillo-Defru). Observations were conducted in the 1UT high-resolution configuration (HR21), using a slow readout mode with $2 \times 1$ detector binning. A simultaneous calibration source was not employed during the observations; instead, the second fiber was used to monitor the sky background. Each spectrum had an exposure time of 40 minutes, yielding a S/N of approximately 30-50 in the reddest spectral orders. 
The reduced spectra were obtained from the ESO Phase 3 Data Release, having been processed with version 2.2.1 of the ESPRESSO DRS. The reduction pipeline performs bias subtraction, flat-fielding, blaze correction, sky subtraction, flux calibration, and spectral extraction. Absolute wavelength calibration is based on exposures of a Thorium-Argon (Th-Ar) hollow cathode lamp. 

Of the initial set of 25 spectra, one was removed because it was taken during a planetary transit and might be affected by the Rossiter-McLaughlin effect. Another spectrum, acquired on June 9, 2019, during a major fiber upgrade of the ESPRESSO spectrograph \citep{Pepe2021}, was also excluded from the analysis. To account for a potential systematic offset introduced by the fiber intervention, we divided the time series into two subsets: 13 RV measurements taken before the upgrade (preUp) and 10 taken after (postUp), resulting in a total of 23 RV data points (Table~\ref{table:rv}). To extract precise RV measurements, we used the SpEctrum Radial Velocity AnaLyser code (\texttt{serval}\footnote{\url{https://github.com/mzechmeister/serval}}), which is based on the template matching technique \citep{Zechmeister2018}. We show the ESPRESSO RV time series in Fig.~\ref{figure:espresso_rv} and provide the corresponding data in Table~\ref{table:espresso_rvs}.

\subsubsection{RV data quality}
\label{sect:rv_data_quality}

We summarize key characteristics of the RV data in Table~\ref{table:rv}. The effective data uncertainty, defined as $\sigma_{\mathrm{eff}} = \left( \frac{1}{N} \sum_{i=1}^{N} \frac{1}{\sigma_i^2} \right)^{-1/2}$, represents the typical per-measurement precision for each dataset, where $N$ is the number of measurements and $\sigma_i$ the formal uncertainty of the $i$-th data point. ESPRESSO demonstrates sub-meter-per-second precision, with 0.53\,m\,s$^{-1}$ before the instrument upgrade and a slightly improved 0.46\,m\,s$^{-1}$ afterward. With 4.35\,m\,s$^{-1}$, CRIRES$^+$ exhibits a larger effective uncertainty, nearly an order of magnitude higher. This difference can be attributed to the lower RV information content in the near-infrared, more conservative uncertainty estimates from the {\tt viper} pipeline, increased telluric contamination, the larger number of model parameters in the RV extraction, and the lower spectral resolving power and instrumental stability relative to ESPRESSO.

The weighted RMS values calculated relative to the weighted mean quantifies the overall scatter in the RV measurements, accounting for individual measurement uncertainties. It captures the excess variability, including contributions from the planetary signal, stellar activity, and instrumental systematics. For ESPRESSO, the weighted RMS is 5.44\,m\,s$^{-1}$ pre-upgrade and 5.96\,m\,s$^{-1}$ post-upgrade. CRIRES$^+$ shows a similar weighted RMS of 5.81\,m\,s$^{-1}$.

\section{Host star properties}
\label{sect:host_star}

\begin{table}
\begin{center}
\caption{Stellar parameters of LHS~3844.}
\label{table:stellar_parameter}      
\begin{tabular}{lll} 
\hline\hline          
\noalign{\smallskip}
Parameter & Value & Reference\tablefootmark{(a)} \\
\noalign{\smallskip}
\hline
\noalign{\smallskip}
\multicolumn{3}{c}{\textit{Basic identifiers}} \\
\textit{Gaia} DR3 ID & 6385548541499112448 &  \\
LHS ID & 3844 & \\
TIC ID & 410153553 & \\
TOI ID & 136 & \\ 
2MASS ID & 22415815-6910089 & \\
\multicolumn{3}{c}{\textit{Coordinates and spectral type}} \\
$\alpha$ [h:m:s] & 22:41:59.12 & \textit{Gaia} DR3 \\
$\delta$ [d:m:s] & $-69$:10:19.95 & \textit{Gaia} DR3 \\
Epoch [yr] & 2016 & \textit{Gaia} DR3 \\
SpT & M4.5 - M5.0 & Van19 \\ 
\multicolumn{3}{c}{\textit{Parallax and kinematics}} \\
$\pi$ [mas] & $67.212 \pm 0.019$ & \textit{Gaia} DR3 \\
$\mu_{\alpha}\cos{\delta}$ [mas\,yr$^{-1}$] & $334.419 \pm 0.021$ & \textit{Gaia} DR3 \\
$\mu_{\delta}$ [mas\,yr$^{-1}$] & $-726.986 \pm 0.020$ & \textit{Gaia} DR3 \\
\multicolumn{3}{c}{\textit{Photometric parameters}} \\
\textit{Gaia} [mag] & $13.365$ & \textit{Gaia} DR3 \\
TESS [mag] & $11.924 \pm 0.008$ & TIC v8.2 \\
$B$ [mag] & $16.942 \pm 0.041$ & TIC v8.2 \\
$V$ [mag] & $15.240 \pm 0.032$ & TIC v8.2 \\
$J$ [mag] & $10.046 \pm 0.023$ & 2MASS \\ 
$H$ [mag] & $9.477 \pm 0.023$ & 2MASS \\
$K$ [mag] & $9.145 \pm 0.023$ & 2MASS \\
\multicolumn{3}{c}{\textit{Stellar parameters}} \\
$M_\star$ [$M_\odot$] & $0.151 \pm 0.014$ & Van19 \\
$R_\star$ [$R_\odot$] & $0.189 \pm 0.006$ & Van19 \\
$L_\star$ [$L_\odot$] & $0.00272 \pm 0.0004$ & Van19 \\
$T_{\rm eff}$ [K] & $3080\pm50$ & This work \\ 
$\log g$ & $5.07\pm0.04$ & This work \\
$\rm{[Fe/H]}$ & $+0.22\pm0.1$ & This work \\
$\rm{[Ti/Fe]}$ & $-0.06\pm0.06$ & This work \\
$P_{\rm rot}$ [d] & $130.0^{+16.9}_{-13.4}$ & This work \\
Age [Gyr] & $7.8\pm1.6$ & Kan20 \\
\noalign{\smallskip}
\hline
\end{tabular}
\tablefoot{
\tablefoottext{a}{
2MASS: \citet{Cutri2003};
\textit{Gaia} DR3: \citet{GaiaDR32023};
Kan20: \citet{Kane2020}
Rai24: \citet{Rains2024};
Van19: \citet{Vanderspek2019};
}}
\end{center}
\end{table}

\subsection{Stellar rotation period}
\label{sect:stellar_rotation_period}

We analyzed the MEarth photometry of LHS~3844 using the generalized Lomb-Scargle (GLS) periodogram \citep{Ferraz-Mello1981, Zechmeister2009} to search for periodic signals. The periodogram reveals a prominent peak corresponding to a period of $P_{\rm rot} = 130.0^{+16.9}_{-13.4}$\,d, suggestive of rotational modulation. We estimated the period uncertainty by fitting a Gaussian to the highest-power peak and adopting its full-width-at-half-maximum (FWHM) as a proxy for the peak width.

Figure~\ref{figure:mearth} shows the light curve, the GLS periodogram, and the phase-folded light curve based on the identified periodicity. A strong and isolated peak in the periodogram is evident, with no significant power at alias or other frequencies. The light curve displays long-term variability that remains consistent across multiple observing seasons. Despite observational gaps, the modulation pattern persists, particularly during later epochs where the temporal sampling is denser. The signal exhibits a coherent, quasi-sinusoidal modulation, which we attribute to rotational variability arising from surface inhomogeneities on the star. The amplitude variations observed over time may reflect the dynamic evolution of these features on the stellar surface. Our findings are consistent with the rotation period reported by \citet{Vanderspek2019}, who derived a value of $128 \pm 24$\,d based on a least-squares periodogram analysis of the same data set.

\begin{figure}
\begin{center}
\includegraphics[width=0.49\textwidth]{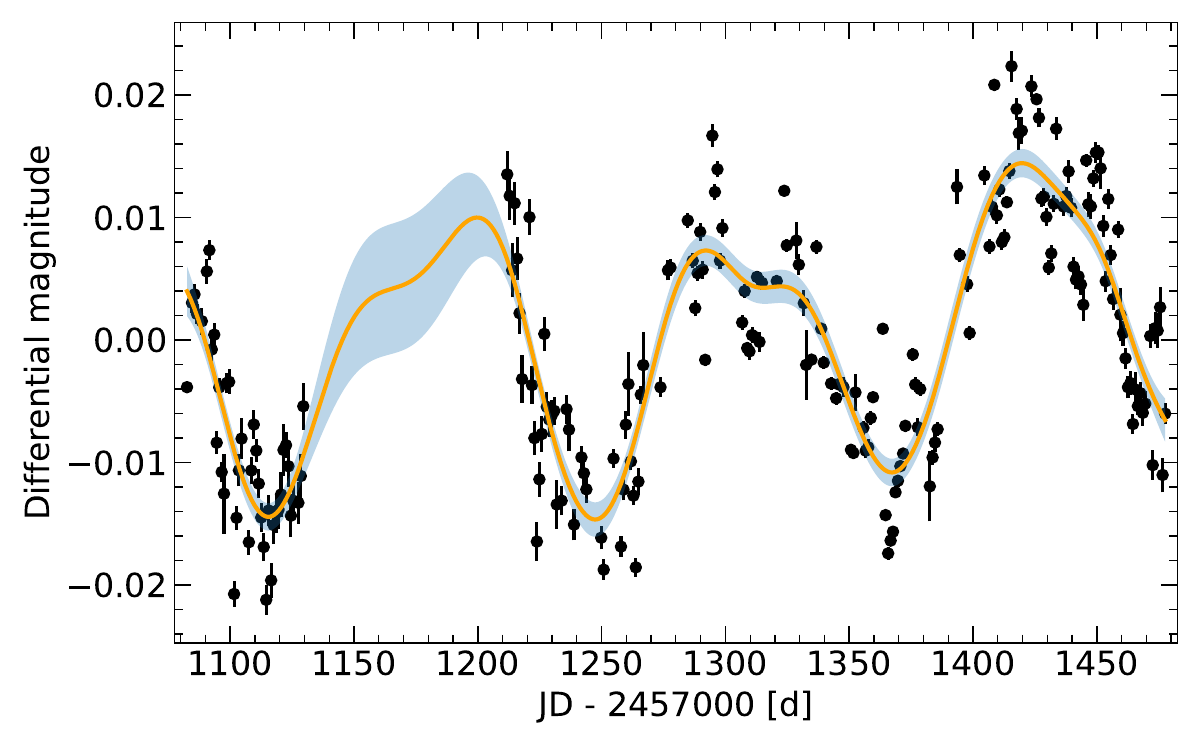}
\caption{\label{figure:mearth_spleaf} 
MEarth photometry of LHS~3844 modeled with a quasi-periodic GP. Black points show the nightly binned MEarth light curve. The solid orange curve denotes the GP predictive mean, and the shaded band indicates the 68\,\% credibility interval.}
\end{center}
\end{figure}

To provide informative constraints for the RV activity modeling (Sects.~\ref{sect:rv_variability}-\ref{sect:independent_rv_analyses}), we modeled the MEarth light curve within a Gaussian Process (GP) framework. A quasi-periodic GP simultaneously captures the rotational timescale and the characteristic evolution time of active regions. We modeled the nightly binned photometry with a squared-exponential periodic (SEP) kernel, restricting the fit to the final portion of the dataset where the quasi-periodic modulation is most apparent and the temporal sampling is densest. The analysis was performed with \texttt{S+LEAF}\footnote{\url{https://www.astro.unige.ch/~delisle/spleaf/doc/index.html}} \citep{Delisle2020, Delisle2022}. The SEP kernel is defined as
\begin{equation}
    k_{i, j}(\tau) = \sigma_{\rm GP, MEarth}^2 \exp \left( -\frac{\tau^2}{2\rho^2_{\rm GP, MEarth}} - \frac{\sin^2(\frac{\pi\tau}{P_{\rm GP, MEarth}})}{2\eta^2_{\rm GP, MEarth}} \right),
\end{equation}
where $\sigma_{\rm GP, MEarth}$ denotes the GP amplitude, $P_{\rm GP, MEarth}$ the characteristic periodic timescale associated with stellar rotation, $\rho_{\rm GP, MEarth}$ the coherence timescale describing the evolution of surface features, $\eta_{\rm GP, MEarth}$ a dimensionless damping parameter controlling the strength of the periodic component, and $\tau = |t_i - t_j|$ is the temporal separation between observations at times $t_i$ and $t_j$. In \texttt{S+LEAF}, the SEP kernel is approximated by a Matérn~3/2 exponential periodic (MEP) kernel, which provides an efficient and accurate representation of quasi-periodic variability. An additional white-noise term, $\sigma_{\rm MEarth}$, was included to account for excess photometric scatter. Broad, weakly informative uniform and log-uniform priors were adopted such that the inferred timescales are driven primarily by the data. The adopted priors and the resulting posterior parameter estimates are summarized in Table~\ref{table:posterior_mearth}, while the posterior distributions of the GP hyperparameters are shown in Fig.~\ref{figure:spleaf_mearth_corner}. The MEarth light curve along with the GP model are shown in Fig.~\ref{figure:mearth_spleaf}. The variability is well reproduced by a smooth, quasi-sinusoidal modulation persisting over multiple seasons, with amplitude variations indicative of evolving active regions. We infer a rotation period of $P_{\rm GP, MEarth} = 124.46^{+14.44}_{-12.12}$\,d, consistent with our GLS-based estimate. The coherence timescale, $\rho_{\rm GP, MEarth} = 103.82^{+137.66}_{-33.90}$\,d, is weakly constrained but suggests active-region lifetimes comparable to one rotation period or longer.

\begin{figure*}
\begin{center}
\includegraphics[width=1.0\textwidth]{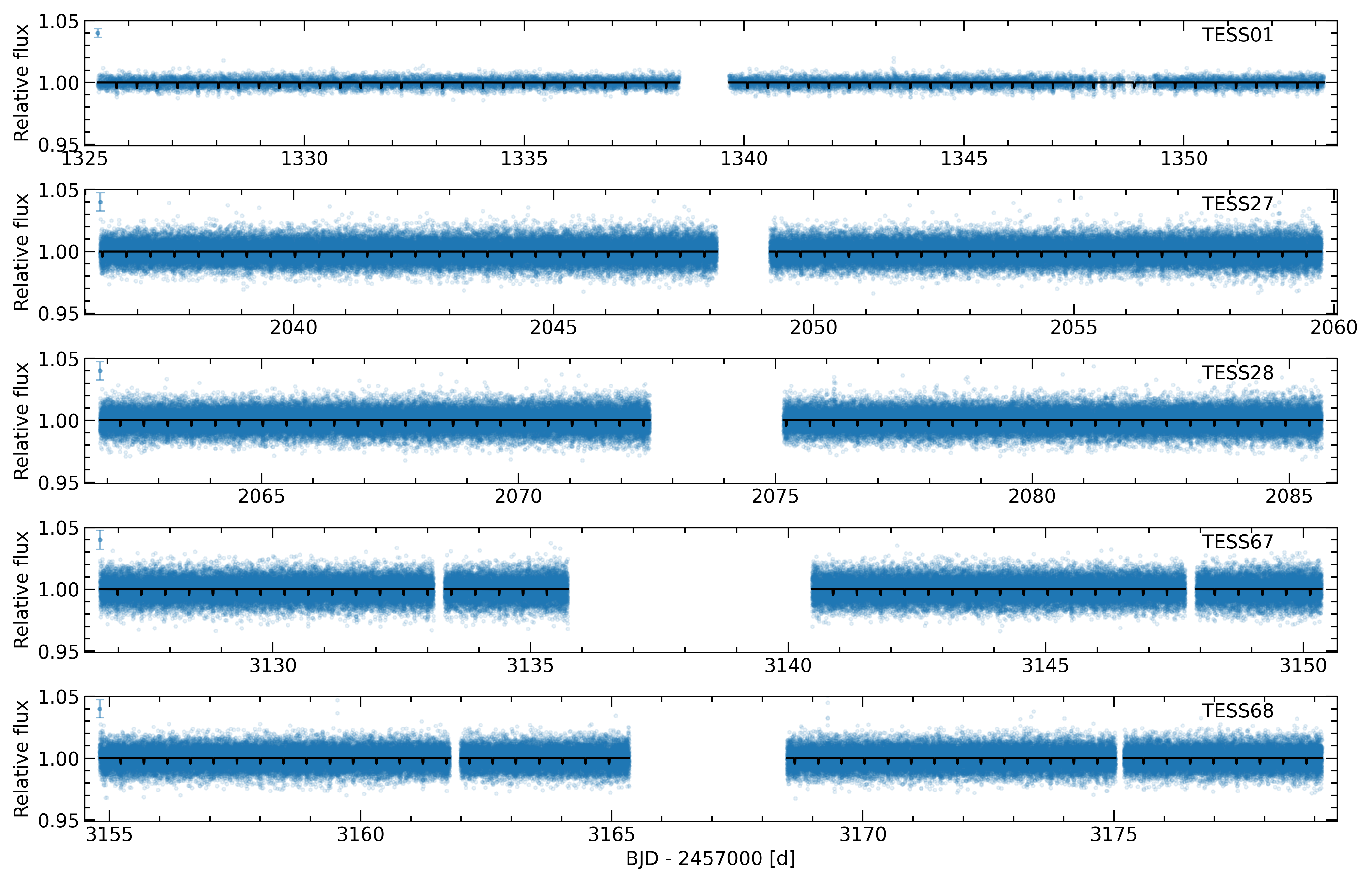}
\caption{\label{figure:juliet_fit_TESS_detrended_transitfit}
Detrended TESS PDCSAP light curve of LHS~3844 for Sectors 1, 27, 28, 67, and 68 (blue dots). 
Overplotted is the best-fit single-planet transiting model (black line). The median photometric uncertainty is illustrated in the upper left corner of each panel.}
\end{center}
\end{figure*}

\subsection{Stellar parameters}

Cool dwarf stars such as LHS~3844 are challenging targets for spectroscopic characterization and chemical abundance analysis. Their cool atmospheres allow for the formation of molecules (e.g., TiO, MgH, and CaH in the optical; H$_2$O, CO, and FeH in the infrared), whose numerous blended absorption features blanket their spectra. This molecular opacity complicates traditional analysis techniques, such as equivalent width measurements or spectral synthesis, even under ideal conditions, particularly when molecular line lists remain incomplete \citep{Hoeijmakers2015, McKemmish2019}. While recent studies have made significant progress in empirically addressing these challenges \citep{Veyette2017, Ishikawa2020, Souto2022}, we adopt alternative, data-driven spectroscopic methods to circumvent these issues.

In recent years, data-driven and statistical frameworks have proven effective for deriving stellar parameters of cool dwarfs \citep{Behmard2019, Birky2020, Maldonado2020, Rains2024, Behmard2025}. Rather than relying on physical models of stellar atmospheres, radiative transfer, and line lists, these methods derive accuracy from libraries of empirically characterized benchmark stars with, for example, temperatures from long-baseline optical interferometry or chemistry from FGK binary companions\footnote{The nuances of data-driven models and cool dwarf benchmarks are beyond the scope of our current work. For a more comprehensive summary, we direct to \citet{Rains2024} and references therein.}. In this work, we employ a version of the data-driven Cannon model \citep{Ness2015}, as implemented in \citet{Rains2024}. The model represents each spectrum as a set of interpretable, per-wavelength polynomial coefficients and is trained on a sample of 103 well-characterized cool dwarf benchmark stars.

We obtained low- to medium-resolution optical spectra of LHS~3844 on August 2019 using the dual-camera Wide-Field Spectrograph \citep[WiFeS;][]{Dopita2007} on the ANU 2.3\,m Telescope at Siding Spring Observatory, NSW, Australia. The dataset includes a blue arm spectrum with a resolution of $\mathcal{R} \sim 3000$ ($350\,\text{nm} < \lambda < 570\,\text{nm}$) and a S/N of $\sim 14$. Additionally, a red arm spectrum is available with a resolution of $\mathcal{R} \sim 7000$ ($540\,\text{nm} < \lambda < 700\,\text{nm}$) and a S/N of $\sim 63$. These observations are part of the surveys published by \citet{Zerjal2021} and \citet{Rains2021}. Both spectra were reduced using the \texttt{PyWiFeS} data reduction pipeline \citep{Childress2014}, with flux calibration as per \citet{Rains2021}. We utilize the published four-parameter model from \citet{Rains2024} to derive the spectroscopic $T_{\rm eff}$, $\log g$, [Fe/H], and [Ti/Fe] for LHS~3844. These values, along with the main characteristics of the star, are summarized in Table~\ref{table:stellar_parameter}.

\section{Analysis and results}
\label{sect:analysis}

For the analysis of TESS photometry and RV measurements, we used {\tt juliet}\footnote{\url{https://juliet.readthedocs.io/en/latest/index.html}} \citep{Espinoza2019}, a Python-based framework designed for the joint modeling of transit light curves and RV data from multiple instruments. {\tt juliet} integrates nested sampling algorithms for robust parameter estimation and enables model comparison through the calculation of Bayesian evidences. Transit models are generated using the {\tt batman} library \citep{Kreidberg2015}, while RV models are produced with {\tt RadVel} \citep{Fulton2018}. For Bayesian inference, we adopted the dynamic nested sampling implementation provided by {\tt DYNESTY} \citep{Speagle2020}, using user-defined prior distributions. This approach efficiently explores the parameter space, yielding posterior samples and the corresponding Bayesian log-evidence, $\ln Z$.

\subsection{Photometric analysis}

\subsubsection{Detrending TESS light curves}
\label{section:detrending}

Visual inspection of the TESS PDCSAP light curves revealed some variability, albeit modest. To mitigate its impact on the determination of planetary parameters, we detrended both the 20-second and the 2-minute cadence PDCSAP light curves from each Sector. This was accomplished using the GP regression library \texttt{celerite} \citep{Foreman-Mackey2017}. We selected an approximated ($\epsilon \to 0$) Matérn-3/2 function as GP kernel as implemented in \texttt{celerite} and integrated into \texttt{juliet},
\begin{equation}
    k_{i, j}(\tau) = \sigma_{\rm GP}^2 \left( 1 + \frac{\sqrt{3} \tau}{\rho_{\rm GP}} \right) \exp \left( -\frac{\sqrt{3}\tau}{\rho_{\rm GP}}\right) ,
\end{equation}
where $\sigma_{\rm GP}$ and $\rho_{\rm GP}$ represent the amplitude and length scale of the GP, respectively, and $\tau = |t_i - t_j|$ is the temporal separation between two data points. To detrend the light curves, we then used the trained GP to predict the model over the full time series, including the transit regions. The original PDCSAP light curves, along with the corresponding GP models, are shown in Fig.~\ref{figure:juliet_fit_TESS_GPDetrending}. For the subsequent analysis, we used the detrended light curves, which are presented in Fig.~\ref{figure:juliet_fit_TESS_detrended_transitfit}.

\subsubsection{Transit signal search}

\begin{figure}
\begin{center}
\includegraphics[width=0.49\textwidth]{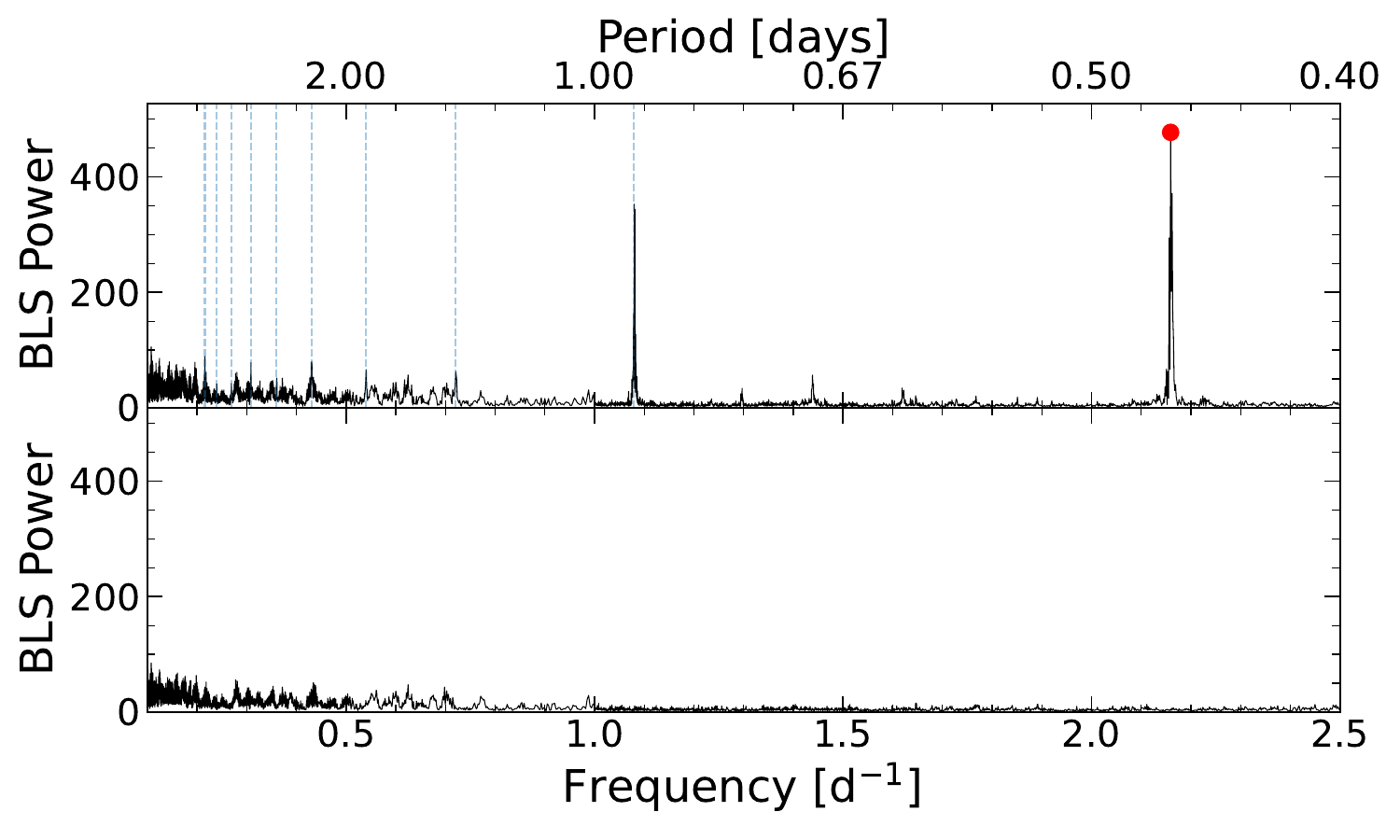}
\caption{\label{figure:bls} 
BLS periodogram of LHS~3844 based on TESS photometry (black lines).
\textit{Top panel:} The red circle marks the highest-power peak, corresponding to the most significant transit signal. Blue dashed lines indicate alias frequencies. 
\textit{Bottom panel:} BLS periodogram after masking transit events caused by LHS~3844~b in the TESS light curve.}
\end{center}
\end{figure}

We used the Box Least Squares (BLS) periodogram \citep{Kovacs2002, Hartman2016}, as implemented in the Python library {\tt astropy.timeseries}\footnote{\url{https://docs.astropy.org/en/stable/timeseries/index.html}}, to verify the transit signal attributed to LHS~3844~b and to search for additional periodic transits in the TESS light curves. The search was conducted over a period range of 0.1 to 100 days, using a finely spaced grid of transit durations motivated by physical considerations.
Figure~\ref{figure:bls} shows the resulting BLS periodogram for the TESS light curve, restricted to the period range up to 10 days for clarity. The strongest peak occurs at a period of 0.463\,d, consistent with the known orbital period of LHS~3844~b. Additional peaks at alias frequencies are also visible. After masking the transits of LHS~3844~b, we reran the periodogram search. No additional significant signals were detected in the masked light curve.

\begin{figure}
\begin{center}
\includegraphics[width=0.49\textwidth]{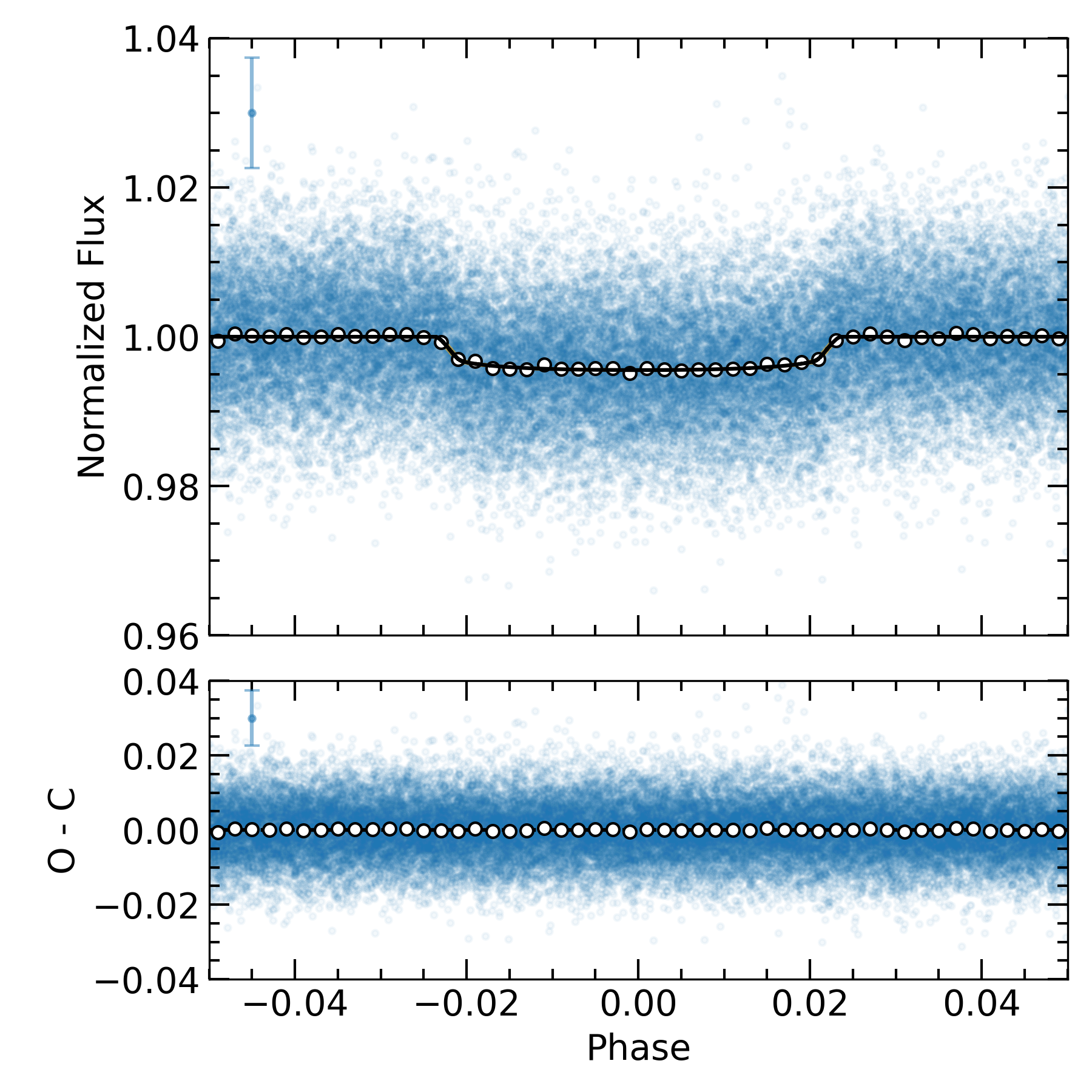}
\caption{\label{figure:tess_phaseFold}
Phase-folded TESS transit light curve of LHS~3844~b. Blue dots are TESS data and white circles are binned data. The best-fit \texttt{juliet} model is overplotted in black. In the bottom panel we show the $\mathrm{O-C}$ residuals. The median photometric uncertainty is illustrated in the upper left corner of both panels.
}
\end{center}
\end{figure}

\subsubsection{Transit-only modeling}
\label{section:transit-only}

As an initial step, we modeled the detrended TESS light curve of LHS~3844~b using {\tt juliet} to refine key transit parameters, particularly the orbital period $P_{\rm b}$ and mid-transit time $t_{0\mathrm{, b}}$. We adopted uniform priors for both parameters.
For $P_{\rm b}$, the prior was centered on the 0.463\,d signal identified by the BLS periodogram, with a conservative range extending $\pm50\,\%$ to account for possible uncertainties. The prior on $t_{0\mathrm{, b}}$ was centered on the most recent observable transit in Sector 68, with a uniform range from 3178.5 to 3179.0\,d ($\mathrm{BJD} - 2\,457\,000$), providing a precise temporal reference for forward propagation of the transit ephemeris. 
Instead of fitting for the scaled semi-major axis $a_{\rm b}/R_\star$, we adopted a parametrization based on the stellar density $\rho_\star$ \citep[e.g.,][]{Espinoza2019}, using a normal prior centered on the value reported by \citet{Vanderspek2019} with a standard deviation corresponding to its $3\sigma$ uncertainty. Rather than sampling directly in the planet-to-star radius ratio and impact parameter, we employed the $(r_1, r_2)$ parametrization introduced by \citet{Espinoza2018}, with uniform priors over the range \mbox{[0, 1]}. For the orbital eccentricity $e$ and argument of periastron $\omega$, we used the parameterization $S_1 = \sqrt{e}\sin{\omega}$ and $S_2 = \sqrt{e}\cos{\omega}$, setting both to zero to enforce a circular orbit.

For the TESS data, we adopted a quadratic limb-darkening law, using the parametrization $q_1$ and $q_2$ from \citet{Kipping2013}, both assigned uniform priors over \mbox{[0, 1]}. We also included a photometric jitter term, modeled with a log-uniform prior between 0 and 1000\,ppm, which is added in quadrature. A relative flux offset was included in the model with a normal prior centered at $M_{\rm TESS} = 0$ and a standard deviation of 0.1. Lastly, the dilution factor was fixed at $D_{\rm TESS} = 1$, under the assumption that the TESS photometry is minimally affected by blending from nearby sources that could mimic a planetary transit signal (Sect.~\ref{sect:tess_photometry}). 

Our analysis yields an orbital period of $P_{\rm b} = 0.462929709^{+0.000000041}_{-0.000000039}$\,d and a mid-transit time of $t_{0\mathrm{, b}} = 3178.833093^{+0.000097}_{-0.000104}$ d for LHS~3844~b. We show the phase-folded TESS light curve as well as the best-fit model in Fig.~\ref{figure:tess_phaseFold}.

\subsection{Spectroscopic analysis}

\subsubsection{Periodogram analysis}
\label{sect:periodogram}

\begin{figure}
\begin{center}
\includegraphics[width=0.49\textwidth]{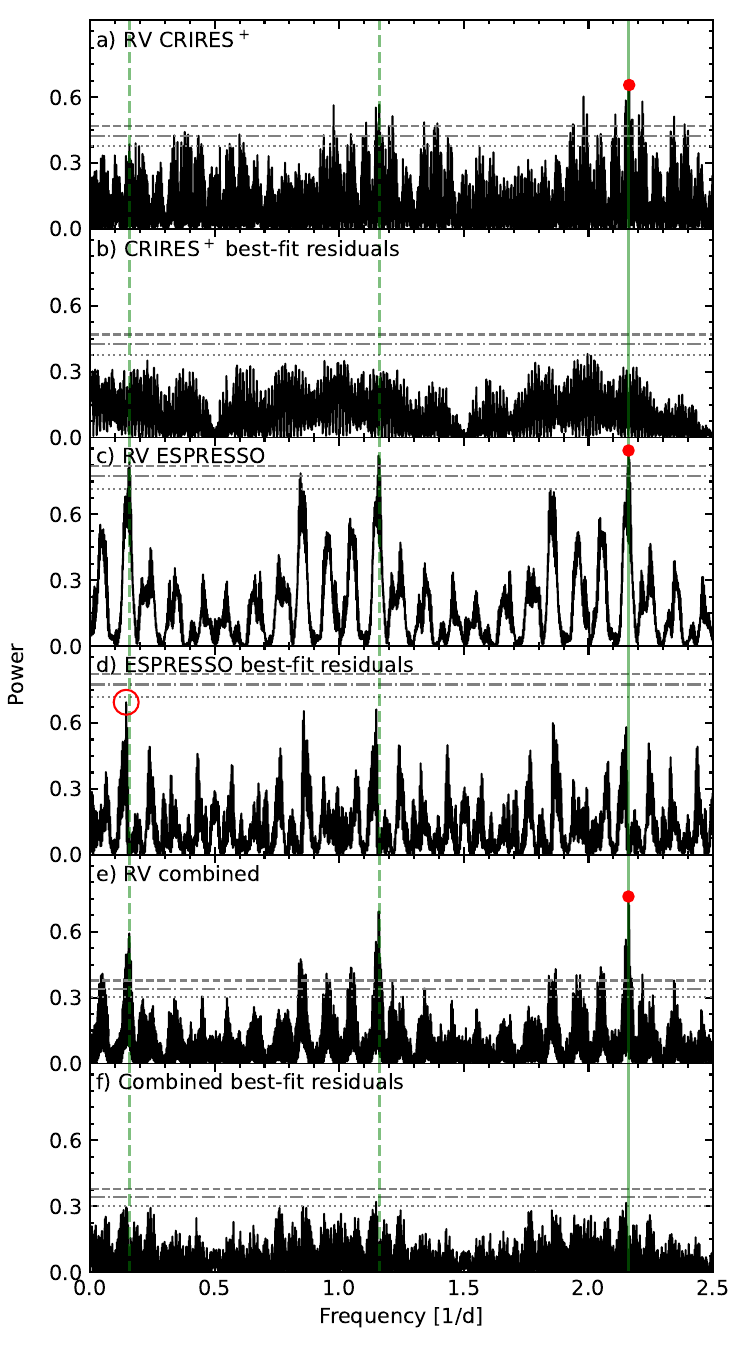}
\caption{\label{figure:periodogram}
GLS periodograms of LHS~3844.
\textit{Panel a:} periodogram for CRIRES$^+$ RVs. Horizontal lines (dashed, dash-dotted, and dotted)
indicate FAP levels at 1\,\%, 0.1\,\%, and 0.01\,\%, respectively. 
The vertical green line highlights the orbital period as determined from the joint fit analysis 
(Sect.~\ref{section:joint}), while the dashed green lines indicate its 1\,d aliases. The highest power peak is marked by a red dot. 
\textit{Panel b:} GLS periodogram for CRIRES$^+$ RV residuals, after subtraction of the Keplerian
model based on the joint fit.
\textit{Panels c-f:} GLS periodograms for ESPRESSO RVs, combined CRIRES$^+$ and ESPRESSO RVs, 
and their best-fit residuals. In \textit{Panel d}, the red circle marks the strongest peak in the periodogram (Sect.~\ref{sect:second_planet}).}
\end{center}
\end{figure}

To investigate the RV variability of LHS~3844, we applied the GLS periodogram to the RVs obtained with CRIRES$^+$ and ESPRESSO. The offsets $\mu_{\rm CRIRES^+}$, $\mu_{\rm ESPRESSO\ preUp}$, and $\mu_{\rm ESPRESSO\ postUp}$, as well as the RV jitter terms $\sigma_{\rm CRIRES^+}$, $\sigma_{\rm ESPRESSO\ preUp}$, and $\sigma_{\rm ESPRESSO\ postUp}$, were adopted from the RV-only analysis, which is described in detail in Sect.~\ref{section:rv-only}. These values were used to shift the RVs on a common zero-point scale and to account for any uncorrelated noise.
Figure~\ref{figure:periodogram} shows the periodograms for the CRIRES$^+$ RVs, ESPRESSO RVs, and their combined datasets. To assess the significance of the peaks in the power spectra, we adopted the normalization described in \citet{Zechmeister2009} and estimated the false alarm probability (FAP) levels of 1\,\%, 0.1\,\%, and 0.01\,\%. 

The periodogram of the CRIRES$^+$ data exhibits a dominant power excess at a frequency of $f = 2.16309017$\,d$^{-1}$ (corresponding to a period of 0.46230158\,d; Fig.~\ref{figure:periodogram}a), with an analytical FAP of $4.6 \times 10^{-8}$\,\%. This frequency is close to the orbital period of 0.46292971\,d derived from the transit-only analysis (Sect.~\ref{section:transit-only}), differing by approximately 0.00063\,d, or about 54 seconds. Additional prominent peaks with FAPs below 0.01\,\% are detected at frequencies consistent with 1\,d aliases of the primary signal. After subtracting the best-fit circular single-planet Keplerian model (1cp; see Sect.~\ref{section:rv-only}), no significant residual power remains above the 1\,\% FAP threshold (Fig.~\ref{figure:periodogram}b). A similar result is obtained for the ESPRESSO dataset (Fig.~\ref{figure:periodogram}c), where the highest peak occurs at 2.16058802\,d$^{-1}$ (0.46283696\,d; analytical FAP $7.2 \times 10^{-5}$\,\%), differing from the transit-only-derived period by approximately 8 seconds. For the combined CRIRES$^+$ and ESPRESSO data (Fig.~\ref{figure:periodogram}e), the strongest peak appears at 2.16019346\,d$^{-1}$ (0.46292150\,d; analytical FAP $2.8 \times 10^{-19}$\,\%), with a difference of $\approx 0.7$ seconds. In both cases, subtraction of the best-fit model eliminates all significant signals from the periodograms, indicating no evidence for additional periodicities (Figs.~\ref{figure:periodogram}d and \ref{figure:periodogram}f, respectively).

\begin{figure}  
\begin{center}
\includegraphics[width=0.49\textwidth]{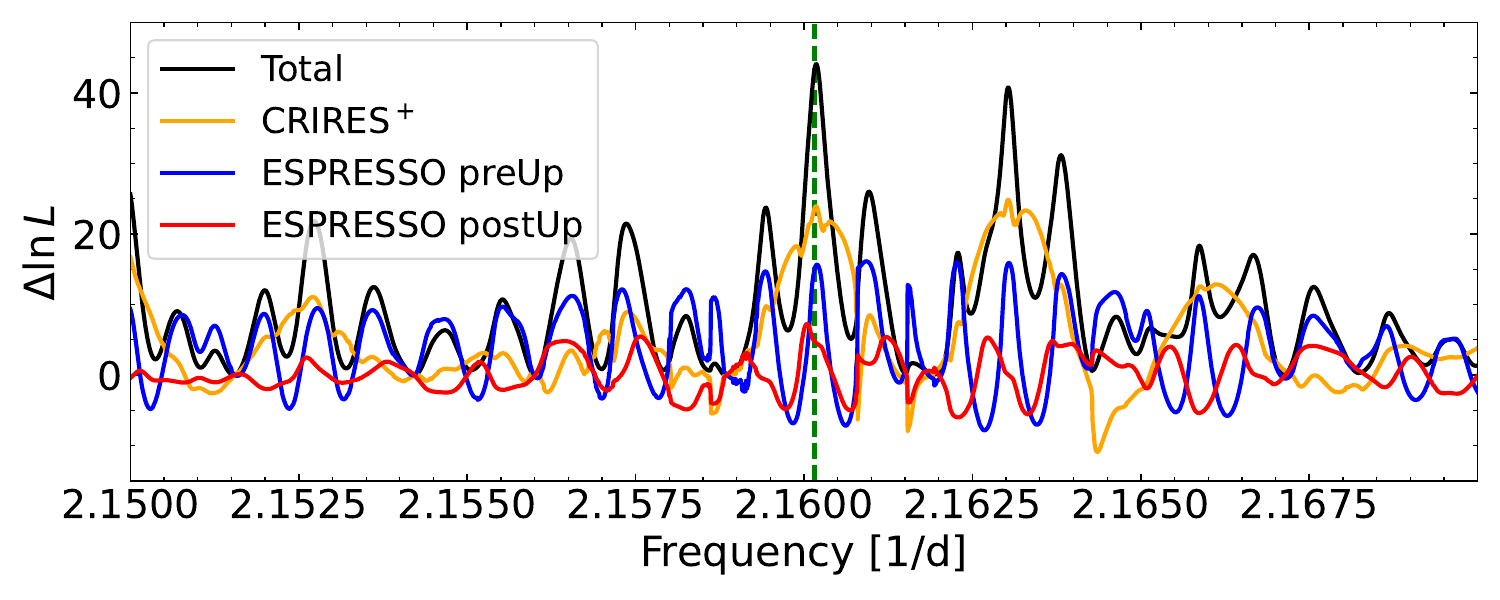}
\caption{\label{figure:mlp}
Maximum-likelihood periodogram of the RV measurements of LHS~3844. The black line shows the periodogram for the combined dataset, while the colored lines correspond to individual instruments (CRIRES$^+$ in orange, ESPRESSO\,preUp in blue, and ESPRESSO\,postUp in red). The vertical dashed line marks the orbital period of LHS~3844~b as determined from TESS photometry.}
\end{center}
\end{figure}

To disentangle the contributions of the different instruments to the detected periodic signal, we applied the maximum-likelihood periodogram\footnote{\url{https://github.com/mzechmeister/python/blob/master/mlp.py}} (MLP) introduced by \citet{Zechmeister2019}. At each trial frequency, the MLP jointly fits a sinusoidal model with common amplitude and phase to all datasets, while allowing instrument-specific offsets and jitters, and reports the relative improvement $\Delta\ln L$ with respect to a constant model. Figure~\ref{figure:mlp} shows the resulting MLP in the frequency range around the orbital period of LHS~3844~b, where the MLP shows a sharp global maximum for the combined dataset. The strongest contribution to the likelihood improvement originates from the CRIRES$^+$ RVs, which dominate the detection and exhibit a peak at the expected orbital frequency. The ESPRESSO subsets also show power at the same frequency, consistent in phase with CRIRES$^+$, but with smaller amplitudes. In particular, the ESPRESSO post-upgrade data yield the weakest contribution. Notably, the sinusoid's amplitude together with the instrument-specific offsets and jitters agree with the parameters from our detailed RV-only modeling (Sect.~\ref{section:rv-only}).

\subsubsection{Spectroscopic activity indicators}

We examined several activity-sensitive spectral diagnostics produced by \texttt{serval} to search for signatures of activity-induced variability. As direct tracers of chromospheric activity, we analyzed the H$\alpha$ and Na D$_1$/Na D$_2$ indices. In addition, we made use of the chromatic index (CRX) and the differential line width (dLW). The CRX \citep{Zechmeister2018} quantifies the wavelength dependence of the RV signal, enabling the identification of spot-induced variations, which typically exhibit a chromatic trend. Rotating starspots can also introduce periodic distortions in the spectral line profiles, which are captured by the dLW indicator. None of the indicators shows a significant peak at the orbital period, nor at its harmonics, and there is likewise no excess power at the stellar rotation period.

\subsubsection{RV-only modeling}
\label{section:rv-only}

We characterized the LHS~3844 system by deriving the orbital parameters of LHS~3844~b using only CRIRES$^+$ and ESPRESSO RVs and no TESS photometry. We began by performing an analysis that remained agnostic of the transit-only results to assess whether the orbital parameters can be independently constrained from the RV data alone. To model the RVs, we used {\tt juliet}, adopting the same priors for $P_{\rm b}$, $t_{0, \mathrm{b}}$, $S_1$, and $S_2$ as those used in the transit-only fit (Sect.~\ref{section:transit-only}). For the semi-amplitude $K_{\rm b}$ we assumed a uniform prior between 0 and 10\,m\,s$^{-1}$. The ESPRESSO observations were split into two separate datasets to account for the instrumental intervention. Together with the CRIRES$^+$ data, this resulted in three instrument-specific RV offsets: $\mu_{\rm CRIRES^+}$, $\mu_{\rm ESPRESSO\ preUp}$, and $\mu_{\rm ESPRESSO\ postUp}$, each assigned a uniform prior between $-100$ and $100$\,m\,s$^{-1}$. Additionally, to account for uncorrelated noise, we included an RV jitter term for each dataset, $\sigma_{\rm CRIRES^+}$, $\sigma_{\rm ESPRESSO\ preUp}$, and $\sigma_{\rm ESPRESSO\ postUp}$, each with a log-uniform prior ranging from 0.001 to 100\,m\,s$^{-1}$. Both the orbital period $P = 0.462313118^{+0.000004629}_{-0.000005321}$\,d and the transit center $t_0 = 3178.828814^{+0.013162}_{-0.014496}$\,d derived from the agnostic analysis are consistent with the ephemerides obtained from the transit-only analysis. While this agreement is robust and serves as a useful sanity check, we incorporated the ephemerides from our transit-only analysis for the subsequent RV-only analysis as these can be measured with significantly higher precision from the transit data. 

We next performed a model comparison analysis via Bayesian evidences focusing on potential orbital eccentricity and the presence of long-term trends indicative of additional long-period companions. To this end, we juxtaposed the performance of four Keplerian models in fitting the RV data: a single planet on a circular orbit (1cp), a single planet on an eccentric orbit (1ep), a single planet on a circular orbit with a linear trend (1cp + lt), and a single planet on an eccentric orbit with a linear trend (1ep + lt). As a null hypothesis, we also consider a model without a planetary signal, fitting only the instrumental parameters under the assumption that the observed variations in the RVs arise purely from instrumental noise (0p). In the eccentric cases, we adopted uniform priors in the range [$-1$, 1] for $S_1$ and $S_2$. To account for any linear trend in the RVs over time, we also included parameters for the intercept $\gamma_0$ and slope $\gamma_1$, assigning them uniform priors over [$-100$, 100]\,m\,s$^{-1}$ and [$-100$, 100]\,m\,s$^{-1}$\,d$^{-1}$, respectively.

The Bayesian evidences of the distinct models are summarized in Table~\ref{tab:bayes_rv-only}. Each model's log-evidence $\ln Z$ is listed, along with the absolute difference relative to the null model ($|\Delta \ln Z|$), which serves as a baseline for comparison. To interpret these values, we adopted the empirical scale proposed by \citet{Trotta2008}, which provides a guideline for assessing the strength of evidence when comparing competing models. These criteria consider a difference of $|\Delta \ln{Z}| = 2.5$ as threshold of moderate evidence (odds of about $12:1$) and $|\Delta \ln{Z}| = 5$ as threshold of strong evidence (odds of about $150:1$).

The null model (0p), which assumes no planetary signal in the RV data, yields $\ln Z = -290.67 \pm 0.41$. This model is decisively disfavored when compared to all models that include at least one planetary companion. The model with the highest Bayesian evidence is the single-planet circular orbit model (1cp), with $\ln Z = -248.72 \pm 0.43$, resulting in $|\Delta \ln Z| = 41.95$ relative to the null hypothesis. Models incorporating additional parameters, either orbital eccentricity or a linear trend, all show reduced Bayesian evidence compared to the 1cp model. The 1cp + lt model yields $|\Delta \ln Z| = 9.15$ less than 1cp, which is substantially disfavored compared to 1cp. Similarly, the eccentric single-planet model (1ep) has a moderate evidence deficit of $|\Delta \ln Z| = 2.89$ relative to the circular orbit model. The model 1ep + lt, which includes both eccentricity and a long-term trend, is the least favored among the one-planet models, with a penalty of $11.48$ in $|\Delta\ln Z|$ compared to 1cp, suggesting that the increase in model complexity is not warranted by the data. In conclusion, the Bayesian evidence supports the 1cp model as the most probable explanation for the observed RV data. There is no compelling statistical justification for including eccentricity or a long-term trend, implying that the signal is best explained by a single planet on a circular orbit around LHS~3844. 

\begin{table}
\centering
\caption{Model comparison using Bayesian evidence from the RV-only analysis of the LHS~3844 system.}
\begin{tabular}{ccc}
\hline\hline
\noalign{\smallskip}
Model\tablefootmark{(a)} & $\ln{Z} $ & $|\Delta \ln{Z}|$ \\
\noalign{\smallskip}
\hline
\noalign{\smallskip}
\multicolumn{3}{c}{\textit{Without stellar activity modeling}} \\
\noalign{\smallskip}
0p & $-290.67\pm0.41$ & 0 \\ 
1cp & $ -248.72\pm0.43$ & $ 41.95$ \\ 
1cp + lt & $ -257.87\pm0.55$ & $ 32.80$ \\
1ep & $ -251.61\pm0.47$ & $ 39.06$ \\ 
1ep + lt & $ -260.19\pm0.57$ & $ 30.48$ \\
\noalign{\smallskip}
\multicolumn{3}{c}{\textit{With stellar activity modeling}} \\
\noalign{\smallskip}
0p + GP & $ -287.85\pm0.41$ & $ 2.82$ \\
\textbf{1cp + GP} & $-247.20\pm0.46$ & $43.47$ \\
1cp + lt + GP & $-256.35\pm0.47$ & $34.32$ \\
1ep + GP & $-249.40\pm0.40$ & $41.27$ \\ 
1ep + lt + GP & $-259.21\pm0.57$ & $31.46$ \\
2cp + GP & $-236.23\pm0.49$ & $54.44$ \\
1ep + 1cp + GP & $-246.21 \pm 0.56$ & $44.45$ \\
\hline
\end{tabular}
\label{tab:bayes_rv-only}
\tablefoot{
\tablefoottext{a}{0p: No planet; 1cp: Circular single-planet Keplerian model; 1cp + lt: Circular single-planet Keplerian model with linear trend; 1ep: Eccentric single-planet Keplerian model; 1ep + lt: Eccentric single-planet Keplerian model with linear trend; GP: Gaussian process with quasi-periodic exponential sine-squared kernel. 2cp: Two planets with circular orbits.}}
\end{table}

\subsection{Stellar activity and additional RV variability}
\label{sect:rv_variability}

Both the MLP analysis (Sect.~\ref{sect:periodogram}) and the RV-only \texttt{juliet} modeling (Sect.~\ref{section:rv-only}) indicate the presence of excess variability beyond the formal internal RV uncertainties, which is captured by the inferred dataset-specific jitter terms. For the preferred circular single-planet model (1cp), we found significant additional noise in the ESPRESSO datasets, with $\sigma_{\rm ESPRESSO\,preUp}=1.72^{+0.48}_{-0.34}$\,m\,s$^{-1}$ and $\sigma_{\rm ESPRESSO\,postUp}=4.06^{+1.19}_{-0.87}$\,m\,s$^{-1}$, while the CRIRES$^+$ time series is consistent with negligible extra jitter, $\sigma_{\rm CRIRES^+}=0.04^{+0.50}_{-0.04}$\,m\,s$^{-1}$. Given that the long-term instrumental stability of ESPRESSO is on the order of $\leq50$\,cm\,s$^{-1}$ \citep{Pepe2021, Schmidt2025, Figueira2025}, we attribute the excess variability primarily to stellar activity or a planetary origin. Both scenarios are examined in the Sects.~\ref{sect:stellar_activity} and \ref{sect:second_planet}.

\subsubsection{Gaussian process modeling of stellar activity}
\label{sect:stellar_activity}

To account for activity-induced RV variability, we introduced a GP noise model. We adopted a quasi-periodic exponential-sine-squared kernel, which is commonly used to model rotationally modulated stellar signals with finite coherence timescales \citep[e.g.,][]{Stock2020, Chaturvedi2022}. In \texttt{juliet}, the kernel is defined as
\begin{equation}
\begin{split}
    k_{i,j}(\tau) &= \sigma_{\rm GP,RV}^2 \exp\!\left(-\alpha_{\rm GP, RV}\,\tau^2 - \Gamma_{\rm GP, RV} \sin^2 \left[ \frac{\pi \tau}{P_{\rm GP, RV}}\right]\right),\\
    \alpha_{\rm GP, RV}&\equiv \frac{1}{2\ell_{\rm GP,RV}^2},
\end{split}
\end{equation}
where $\tau = |t_i - t_j|$, $\sigma_{\rm GP, RV}$ is the GP amplitude, $\ell_{\rm GP,RV}$ is the coherence timescale, $\Gamma_{\rm GP, RV}$ controls the strength of the periodic modulation, and $P_{\rm GP, RV}$ the characteristic recurrence period associated with stellar rotation.\\
When adopting uninformative priors, the RV data alone do not meaningfully constrain the quasi-periodic hyperparameters because the observations sample only a limited fraction of the stellar rotation cycle ($P_{\rm rot}\approx130\,$d). The ESPRESSO pre- and post-upgrade time series cover only $\sim 25\,\%$ (32\,d) and $\sim 45\,\%$ (58\,d) of $P_{\rm rot}$, respectively, yielding insufficient phase coverage to robustly identify the quasi-periodic kernel hyperparameters. We therefore imposed informative Gaussian priors on $\alpha_{\rm GP,RV}$, $\Gamma_{\rm GP,RV}$, and $P_{\rm GP,RV}$ based on the independent GP analysis of the MEarth photometry presented in Sect.~\ref{sect:stellar_rotation_period}. The corresponding priors are listed in Table~\ref{table:posterior}. We note that the quasi-periodic GP kernel adopted here differs in functional form from the squared-exponential periodic (SEP/MEP) kernel used to model the MEarth photometry. While both kernels describe quasi-periodic variability associated with stellar rotation, their hyperparameters are not equivalent. In particular, the coherence timescale inferred from the photometric SEP/MEP kernel reflects the evolution of active regions and does not map one-to-one onto the characteristic decay timescale of the quasi-periodic kernel used in the RV analysis. Therefore, when using the MEarth-based constraints to inform the RV GP priors, the correspondence between hyperparameters is only approximate.

We repeated the Bayesian model comparison including the GP component and summarize the resulting evidences in Table~\ref{tab:bayes_rv-only}. The model ranking remains unchanged, with the circular single-planet model still strongly favored. Including the GP yields a modest improvement in the Bayesian evidence relative to the corresponding white-noise model (e.g., $|\Delta \ln Z| = 1.52$ for 1cp + GP compared to 1cp), indicating that correlated noise is weakly preferred but not dominant. Despite the modest statistical improvement, we choose 1cp + GP as our fiducial model because the GP provides a physically motivated treatment of activity-driven correlated noise and leaves the inferred planetary parameters unchanged.

With the GP included, the inferred jitter for the ESPRESSO pre-upgrade data remains essentially unchanged at $\sigma_{\rm ESPRESSO\,preUp}=1.70^{+0.49}_{-0.37}$\,m\,s$^{-1}$. In contrast, the post-upgrade jitter is reduced to $\sigma_{\rm ESPRESSO\,postUp}=2.39^{+1.15}_{-0.72}$\,m\,s$^{-1}$. Several recent studies have demonstrated that ESPRESSO can achieve residual RV scatter at the level of $\sim 0.5$\,m\,s$^{-1}$ for slowly rotating M dwarfs \citep[e.g.,][]{Faria2022, GonzalezHernandez2024}. In this context, the observed excess variability likely exceeds the nominal instrumental noise floor and points to a combination of stellar variability and additional, potentially coherent signals not captured by the current single-planet model, as further explored in the next section.

\subsubsection{Search for additional planetary signals}
\label{sect:second_planet}

Although the quasi-periodic GP accounts for a fraction of the excess variability, non-negligible jitter remains in the ESPRESSO data under the preferred 1cp + GP model. We therefore investigated whether an additional Keplerian signal could be present in the RV time series. \\
We first inspected the GLS periodogram of the ESPRESSO residuals after subtraction of the best-fit model. The strongest remaining peak occurs at a period of 6.887\,d, with an analytical FAP of 2.1\,\%, which is not formally statistically significant (red circle in Fig.~\ref{figure:periodogram}d). 
To assess whether a coherent Keplerian signal at this period is nevertheless supported in a global framework, we performed a two-planet fit with \texttt{juliet}. The model consists of three components: two planets on circular orbits and the same constrained quasi-periodic GP. The priors for the inner planet and the GP were identical to those adopted in the 1cp + GP analysis.
In an exploratory search, we adopted a log-uniform prior on the period of the outer companion between 2 and 1000\,d. The posterior distribution converged toward the solution near $6.88\,$d. We therefore refined the analysis using a uniform prior on $P_{\rm c}$ between 2 and 10\,d, a uniform prior on $t_{0,\rm c}$ spanning $t_{0,\rm b}$ to $t_{0,\rm b}+10\,$d, and a uniform prior on the semi-amplitude $K_{\rm c}$ between 0 and 10\,m\,s$^{-1}$. The orbit was assumed to be circular.
The resulting Bayesian evidence for the 2cp + GP model is $\ln Z = -236.23 \pm 0.49$ (Table~\ref{tab:bayes_rv-only}), corresponding to $|\Delta \ln Z| = 10.97$ relative to the 1cp + GP model, which constitutes strong evidence in favor of the more complex model. The improvement in evidence is accompanied by a reduction of the ESPRESSO jitter terms to $\sigma_{\rm ESPRESSO\,preUp}=0.14^{+0.27}_{-0.12}\,$m\,s$^{-1}$ and $\sigma_{\rm ESPRESSO\,postUp}=0.02^{+0.13}_{-0.01}\,$m\,s$^{-1}$, while the GP amplitude and the parameters of LHS~3844~b remain consistent with the 1cp + GP solution. For the candidate companion, we obtain an orbital period of $P_{\rm c}=6.8758^{+0.0057}_{-0.0035}$ and a semi-amplitude of $K_{\rm c}=3.05^{+0.25}_{-0.26}\,$m\,s$^{-1}$, corresponding to a minimum mass of $m_{\rm c}\sin{i}=2.56^{+0.27}_{-0.26}\,M_\oplus$. After subtraction of the full 2cp + GP model, the weighted RMS of the ESPRESSO residuals decreases to 0.45\,m\,s$^{-1}$ (pre-upgrade) and 0.30\,m\,s$^{-1}$ (post-upgrade), consistent with the instrumental precision.
The adopted priors and resulting posteriors are summarized in Table~\ref{table:posterior_second_planet}. Figures~\ref{figure:espresso_rv_two_planets}-\ref{figure:corner_rv_only_two_planets} illustrate the time series, phase-folded signals, and posterior correlations for the two-planet solution.\\
We note that the candidate signal at 6.88\,d lies close to the expected second-order daily alias of LHS~3844~b at $\sim 6.244\,$d (Fig.~\ref{figure:periodogram}d). Imperfect subtraction of the inner-planet, particularly if its orbit is eccentric while being modeled with a circular Keplerian, can leave residual power at the true frequency, which the spectral window redistributes into the corresponding alias frequencies \citep{Dawson2010}. To test this scenario, we performed an additional fit allowing the inner planet to have a nonzero eccentricity while retaining a single circular companion model for the outer signal (1ep + 1cp + GP; Table~\ref{tab:bayes_rv-only}). The inferred eccentricity is $e_{\rm b}=0.02^{+0.03}_{-0.02}$, consistent with a circular orbit within uncertainties. The outer Keplerian component near 6.88\,d remains present in the fit with consistent parameters and amplitude, indicating that the signal does not disappear when eccentricity of the inner planet is allowed. Allowing for a small eccentricity of the inner planet therefore does not remove the residual signal near 6.88\,d, and an additional Keplerian component remains a plausible explanation for the excess variability in the ESPRESSO RVs.\\
Although USP planets are frequently found in compact multi-planet systems \citep{Winn2018} and despite the formally strong Bayesian preference for the 2cp + GP model, we refrain from claiming the detection of a second planet as the signal is not independently significant in the residual periodograms. Additional dedicated RV monitoring with extended time coverage is required to confirm or reject the planetary interpretation of the 6.88\,d signal. We searched the TESS light curves for transits and found no significant BLS peak near this period.

\subsection{Joint analysis}
\label{section:joint}

To determine the most precise planetary parameters of the LHS~3844 system, we performed a joint fit to the TESS photometry and the CRIRES$^+$ and ESPRESSO RV data. The final Keplerian model consists of a single planet on a circular orbit and a GP. For the joint analysis, we adopted the same prior choices as in the transit-only and RV-only fits. The priors used in the final joint fit are listed in Table~\ref{table:posterior}.

\begin{figure}
\begin{center}
\includegraphics[width=0.49\textwidth]{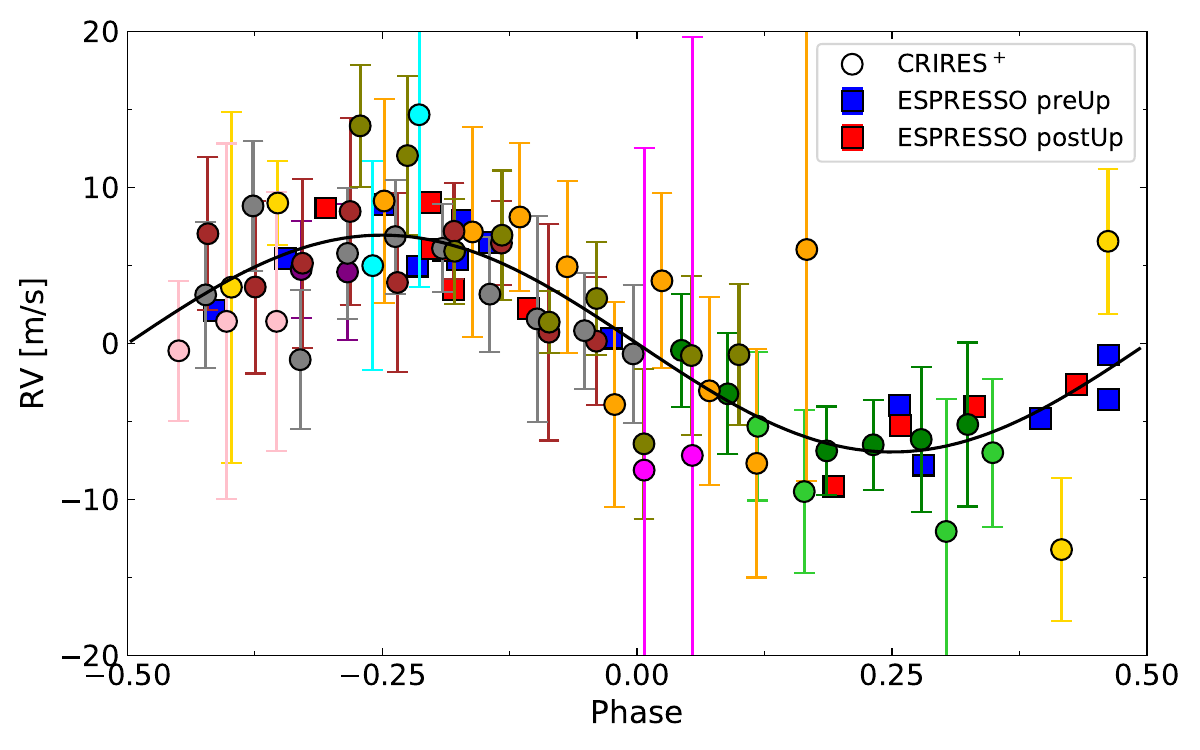}
\caption{\label{figure:rv_phasefold}
Phase-folded RV measurements of LHS~3844, shown alongside the best-fit Keplerian model for a single planet on a circular orbit (black curve). The colors and shapes of the data points correspond to those depicted in Figs.~\ref{figure:crires_rv} and \ref{figure:espresso_rv}. The error bars of the ESPRESSO data are smaller than the plotted symbols.}
\end{center}
\end{figure}

Adopting the circular, single-planet model, we measure a semi-amplitude of $K_{\rm b}=6.95^{+0.55}_{-0.60}$\,m\,s$^{-1}$ and an orbital period of $P_{\rm b}=0.462929709^{+0.000000044}_{-0.000000042}$\,d. Combined with the planetary radius of $R_{\rm b} = 1.286^{+0.043}_{-0.044}\,R_\oplus$, these values yield a planet mass of $m_{\rm b} = 2.37\pm0.25\,M_\oplus$ and a bulk density of $\rho_{\rm b} = 6.15^{+0.60}_{-0.61}$\,g\,cm$^{-3}$. The best-fit model is shown together with the CRIRES$^+$ data in Fig.~\ref{figure:crires_rv} and the ESPRESSO data in Fig.~\ref{figure:espresso_rv}. The weighted RMS of the residuals is $3.05$\,m\,s$^{-1}$ for CRIRES$^+$, $1.52$\,m\,s$^{-1}$ for ESPRESSO preUp, and $1.99$\,m\,s$^{-1}$ for ESPRESSO postUp (Table~\ref{table:rv}). The phase-folded RV time series is shown in Fig.~\ref{figure:rv_phasefold}. The best-fit Keplerian orbital parameters are reported in Table~\ref{table:posterior}, and the corresponding posterior distributions are shown in Fig.~\ref{figure:tess_phaseFoldet_corner}.

\subsection{Independent CRIRES$^+$ and ESPRESSO RV analyses}
\label{sect:independent_rv_analyses}

To assess the ability of CRIRES$^+$ to independently recover the Doppler signal of LHS~3844~b, we performed separate RV-only analyses of the CRIRES$^+$ and ESPRESSO datasets. For each instrument, we repeated the model comparison described in Sect.~\ref{section:rv-only}, testing circular and eccentric single-planet models with and without a linear trend, as well as GP noise terms. The corresponding Bayesian evidences are listed in Table~\ref{tab:bayes_rv-only_crires}, and posterior estimates for the preferred models appear in Table~\ref{table:posterior_only_crires}.

For CRIRES$^{+}$, the circular single-planet model (1cp) is decisively favored over the no-planet model ($|\Delta \ln Z| = 23.97$). Adding a GP yields only a marginal and statistically insignificant increase in the evidence ($|\Delta \ln Z| = 0.69$), well below the threshold at which increased model complexity would be justified. Correspondingly, the GP amplitude remains poorly constrained ($\sigma_{\rm GP,RV}=1.39^{+1.61}_{-1.39}$\,m\,s$^{-1}$), and the instrumental jitter term is consistent with zero.
The recovered semi-amplitude is stable across models: $K_{\rm b}=6.84^{+0.82}_{-0.83}$\,m\,s$^{-1}$ for the 1cp model and $K_{\rm b}=7.19^{+1.00}_{-0.94}$\,m\,s$^{-1}$ when including a GP. Both values are fully consistent with the joint CRIRES$^+$ \& ESPRESSO result (Sect.~\ref{section:joint}) and with the transit-only ephemeris (Sect.~\ref{section:transit-only}). Thus, CRIRES$^{+}$ alone robustly recovers the planetary signal at the correct period and amplitude without requiring a correlated-noise model.

For ESPRESSO, the RV-only model comparison provides a useful independent cross-check. The 1cp model is strongly favored over the no-planet hypothesis ($|\Delta \ln{Z}|=16.47$), while adding eccentricity, a trend, or a GP term is not supported by the evidence ($|\Delta \ln{Z}|=0.24$ for 1cp + GP relative to 1cp). The inferred semi-amplitude $K_{\rm b}=7.03^{+0.91}_{-0.90}\,$m\,s$^{-1}$ agrees with both the CRIRES$^+$-only recovery and the joint fit. Together, these results provide an external validation that the CRIRES$^+$ analysis retrieves the same signal at the correct period and amplitude.

\section{Discussion}

\subsection{Comparison to USP population}

\begin{figure}
\begin{center}
\includegraphics[width=0.49\textwidth]{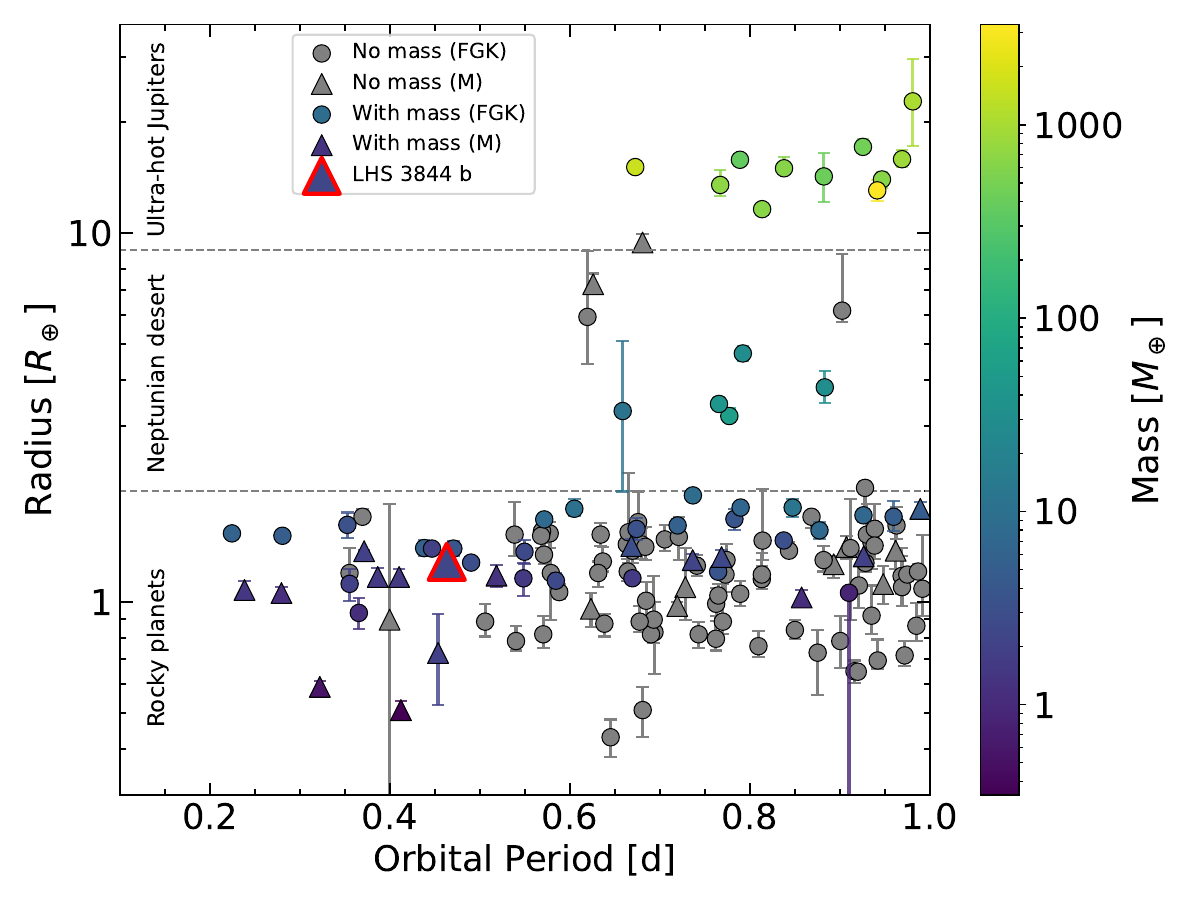}
\caption{\label{figure:radius_vs_period}
Radius versus orbital period for transiting USP planets. Planets orbiting around FGK host stars are shown as circles, and those around M stars are shown as triangles. Planet masses are shown color-coded. Planets without mass measurements are plotted in gray. LHS~3844~b is highlighted with a red border. Horizontal dashed lines indicate three radius regimes: rocky planets, the Neptunian desert, and ultra-hot Jupiters.}
\end{center}
\end{figure}

With its key properties established, we contextualize LHS~3844~b within the current population of known USP planets. In Fig.~\ref{figure:radius_vs_period}, we show its location in the period-radius plane alongside other USP planets. We identify three distinct radius regimes among USP planets: planets presumed to be rocky with radii $\lesssim\,$2\,$R_\oplus$, the Neptunian desert, encompassing planets with radii $2\,R_\oplus \lesssim R \lesssim 9\,R_\oplus$ \citep{Mazeh2016}, and ultra-hot Jupiters with radii exceeding $\gtrsim9\,R_\oplus$. Although the USP population spans this full range, it is dominated by small rocky worlds, with only a handful of Jupiter-sized objects and a sparsely populated sub-Jovian regime. LHS~3844~b lies firmly in the rocky regime, well below the lower boundary of the Neptunian desert, which can be explained by the photoevaporation of highly irradiated sub-Neptunes, and thus supports the interpretation that LHS~3844~b is a bare, rocky core, likely having lost any primordial atmosphere due to intense stellar irradiation. Its placement in a densely populated region of the period-radius plane, alongside other well-characterized USP planets with measured masses, suggests it is representative of a broader population shaped by similar evolutionary pathways. As such, LHS~3844~b serves as a benchmark for studying atmospheric erosion and the interior structure of short-period terrestrial exoplanets.

\subsection{Planetary composition and core-mass fraction}

\begin{figure}
\begin{center}
\includegraphics[width=0.49\textwidth]{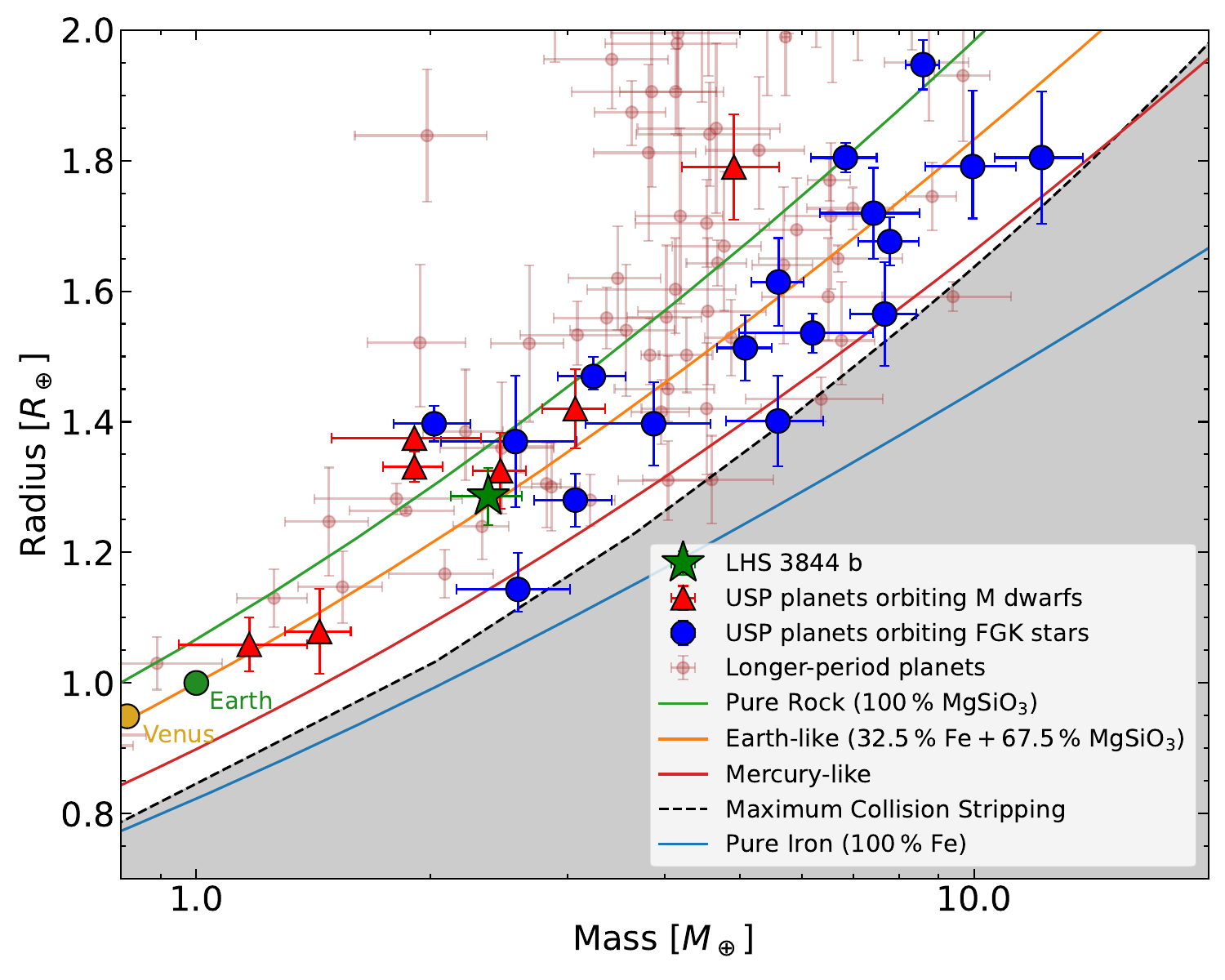}
\caption{\label{figure:mass_radius}
Mass-radius diagram for rocky exoplanets with ultra-short ($P < 1$\,d) and longer orbital periods. The colored points show planets with well-constrained masses and radii, separated by host star type and orbital period: USP planets orbiting M dwarfs (red), USP planets orbiting FGK stars (blue), and longer-period planets orbiting FGKM stars (brown).
LHS~3844~b is highlighted as a green star. Earth and Venus are shown for comparison.
Solid curves denote theoretical mass-radius relations from \citet{Zeng2019} for pure iron (blue), Earth-like composition (orange), and pure rock (green), and from \citet{Brugger2017} for a Mercury-like composition.
The dashed black line marks the maximum collisional stripping limit from \citet{Marcus2010}.}
\end{center}
\end{figure}

Observational evidence is consistent with LHS~3844~b lacking a substantial atmosphere and being most likely a bare-rock planet, although a tenuous, replenished atmosphere cannot be entirely ruled out \citep{Kreidberg2019, Diamond-Lowe2020}. In the absence of a significant atmosphere, compositional degeneracies are reduced \citep{Rogers2010}, and the precise measurements of the planet's mass ($\approx 11\,\%$) and radius ($\approx 3\,\%$) enable robust constraints on its bulk and core composition. In Fig.~\ref{figure:mass_radius}, we show the position of LHS~3844~b on a mass-radius diagram alongside the known USP and Earth-sized planets whose masses and radii are measured to better than $25\,\%$ and $8\,\%$, respectively\footnote{Data from \url{https://exoplanet.eu/}.}. These thresholds were chosen so that mass and radius contribute comparably to the propagated uncertainty in bulk density \citep{Otegi2020}. We further include iso-composition models from \citet{Zeng2016, Zeng2019} and \citet{Brugger2017}, as well as the lower envelope of the mass-radius relation for super-Earths, representing the maximum iron content allowed by models of collisional mantle stripping \citep{Marcus2010}. The positions of Venus and Earth are also shown for comparison. Of the 25 USP planets that satisfy our threshold, eight, including LHS~3844~b, orbit M stars. The position of LHS~3844~b in the mass-radius diagram suggests a bulk density consistent with an Earth-like composition.

Using the measured mass and radius, we estimated the core mass fraction (CMF), which is the fraction of a planet's total mass that resides in its core, to further constrain the interior composition of LHS~3844~b. Because of the absence of a volatile envelope or water layer, the CMF is more accurate in this case and not a lower limit. We first applied the semi-empirical mass-radius relation for two-layer rocky planets \citep{Zeng2016} based on the Preliminary Reference Earth Model \citep[PREM;][]{Dziewonski1981}. This relation is solely parameterized by the CMF. 
Within this framework, the CMF for LHS~3844~b is $0.25\pm0.22$. However, this value is statistically consistent with zero at $\sim 1\sigma$ level and we therefore interpret this result as an upper limit. The inferred CMF is consistent with a wide range of rocky interior structures, including an Earth-like composition ($\textrm{CMF}\approx0.33$) and iron-poor interiors. In this simple two-layer model with a pure-iron core and silicate-rock mantle, the CMF is equal to the iron mass fraction (Fe-MF).

We next used the \texttt{exopie} code \citep{Plotnykov2024}, which builds upon the interior structure model \texttt{SuperEarth} \citep{Valencia2006, Plotnykov2020}. Unlike the simpler bi-layer (core-mantle) structure, this model adopts a differentiated two-layer architecture that resolves both layers in greater detail. The core is partitioned into a liquid outer core and a solid inner core, with the phase boundary set by the melting curve of an iron alloy. The mantle is stratified into four Earth-analog layers (upper mantle, transition zone, lower mantle, and lowermost mantle) separated by pressure- and temperature-dependent phase transitions. In addition to the CMF, \texttt{SuperEarth} reports an Fe-MF that accounts for iron partitioning into the silicate mantle, while light elements may be present in the iron core. In our analysis, we considered varying degrees of core differentiation, parameterized by the iron fraction in the mantle x$_{\rm Fe}$, as well as varying silicon abundances in the core x$_{\rm Si}$. In order to solve for the composition given the mass, radius, and the uncertainties, \texttt{exopie} provides an MCMC framework that is coupled to the interior structure model. We adopted skewed normal priors for the mass and radius and uniform priors informed by terrestrial planet compositions, $\mathrm{x_{Fe}} \sim U(0,20)\,\%$ by mol and $\mathrm{x_{Si}} \sim U(0,20)\,\%$ by mol. For LHS~3844~b we find a CMF of $0.26^{+0.17}_{-0.16}$ and an Fe-MF of $0.26^{+0.13}_{-0.12}$. As in the simpler two-layer analysis, the posterior distribution is broad and remains consistent with zero at $<2\sigma$ significance. Consequently, the inferred CMF should be interpreted as a weak constraint or upper limit. The posterior distributions of the fitted parameters are shown in Fig.~\ref{figure:exopie}. The CMF value from the \texttt{exopie} analysis is consistent with those obtained with the relation from \citet{Zeng2016}. 

\subsection{Evolutionary implications}

Coupled outgassing-erosion models by \citet{Kane2020} delineate possible pathways for the observed atmospheric paucity, including removal by a single giant impact. In particular, they show that an impactor of $0.26$-$0.35\,M_\oplus$ striking at roughly the orbital velocity of LHS~3844~b could eject a massive secondary atmosphere, but would almost certainly entail severe mantle stripping and yield an Fe-rich super-Mercury with CMF $\sim70\,\%$ \citep{Marcus2009, Marcus2010, Bonomo2019}. Our results on the CMF, however, is compatible with a nominally Earth-like interior and inconsistent with the mantle-stripping outcome required by the giant-impact hypothesis. We therefore reject the giant-impact explanation for LHS~3844~b's lack of atmosphere.

\citet{Kane2020} further argue that the absence of a detectable atmosphere naturally follows if the planet formed volatile-poor interior to the water snow line, and they note that such a volatile-poor origin could reflect the overall system configuration and, specifically, the presence of an unobserved massive outer companion that restricted volatile delivery and/or shaped early migration. Our multi-year RV timeseries provides no support for such a body. The Bayesian model comparison decisively favors a single, circular-orbit model for LHS~3844~b over models that include a long-term acceleration, and the GLS periodograms of the individual and combined RVs, along with the residuals after subtracting the best-fit Keplerian, show no additional significant signals. In concert with the lack of any trend in the RVs, these results indicate no observational evidence for additional companions in the system, thereby disfavoring the ``hidden massive companion'' pathway to a volatile-poor composition proposed by \citet{Kane2020}. 

\subsection{Host-star chemistry and planetary composition}

Beyond dynamical and atmospheric evolution, a complete understanding of rocky exoplanets such as LHS~3844~b ultimately requires connecting their bulk and interior properties to the chemistry of their host stars. Stellar abundance ratios, particularly Fe/Mg/Si, C/O, and refractory-to-volatile trends, set the initial composition of planet-forming material and thus influence planetary core-mantle fractions and densities \citep[e.g.,][]{Bond2010, Thiabaud2014, Thiabaud2015, Bitsch2020, Spaargaren2023}. It has been suggested that variations in stellar chemistry may produce systematic differences in planet bulk density and internal structure \citep[e.g.,][]{Dorn2019, Adibekyan2021, Adibekyan2024, Behmard2025_b}. In this context, stripped planets such as LHS~3844~b serve as valuable laboratories for studying planetary composition, since their rocky interiors are directly accessible through mass-radius measurements. Linking such measurements to host-star abundances is therefore a key step toward a chemical framework for planetary diversity. This field remains in its infancy, but future surveys combining precise stellar spectroscopy with density measurements of rocky planets will clarify whether composition-dependent trends emerge across the exoplanet population.

\section{Summary}

We have measured the mass of the USP super-Earth LHS~3844~b using a near-infrared RV data set from CRIRES$^+$ with a $K$-band gas cell, complemented by archival ESPRESSO spectroscopy and a joint fit with TESS photometry. 
The preferred model is a single planet on a circular orbit and a GP to account for stellar activity. The final joint fit (1cp + GP) yields an orbital period of $P_{\rm b} = 0.462929709^{+0.000000044}_{-0.000000042}\,$d, an RV semi-amplitude of $K_{\rm b} = 6.95^{+0.55}_{-0.60}\,$m\,s$^{-1}$, and hence a planet mass of $m_{\rm b} = 2.37\pm0.25\,M_\oplus$ for a transit radius $R_{\rm b} = 1.286^{+0.043}_{-0.044}\,R_\oplus$. The resulting bulk density $\rho_{\rm b} = 6.15^{+0.60}_{-0.61}\,$g\,cm$^{-3}$ is consistent with a predominantly rocky interior. 

In RV-only Bayesian model comparisons, we found no compelling evidence for eccentricity or a long-term trend. We introduced a GP to model stellar activity, motivated by excess variability absorbed into the ESPRESSO jitter terms. 
Including a quasi-periodic GP is at most weakly preferred and does not meaningfully change the inferred planetary parameters. Overall, the remaining excess variability likely exceeds the nominal instrumental noise floor and instead suggests contributions from stellar variability and possible additional planetary signals. In particular, after subtracting the preferred one-planet model, the ESPRESSO residual periodogram shows a signal at 6.88\,d with $\mathrm{FAP}\approx 2.1\,\%$. Although this peak is not formally significant, Bayesian model comparison favors a two-planet model with a GP, suggesting that an additional coherent RV component may be present and motivating further RV monitoring.

Our instrument-specific RV analyses further show that the planet is independently detected in both CRIRES$^{+}$ and ESPRESSO. For both instruments, the detection is significant without requiring GP modeling and the inferred semi-amplitude is stable across models. A CRIRES$^+$-only fit yields $K_{\rm b} = 6.84^{+0.82}_{-0.83}$\,m\,s$^{-1}$ ($K_{\rm b} = 7.19^{+1.00}_{-0.94}$\,m\,s$^{-1}$ when including a GP), and an ESPRESSO-only fit gives $K_{\rm b} = 6.78^{+0.65}_{-0.64}$\,m\,s$^{-1}$ ($K_{\rm b} = 6.75^{+0.68}_{-0.65}$\,m\,s$^{-1}$ when including a GP), all consistent with the joint solution.

Interior modeling places only a weak constraint on the iron-core mass fraction, with $\mathrm{CMF}=0.25\pm0.22$ from semi-empirical mass-radius relations, and $\mathrm{CMF}=0.26^{+0.17}_{-0.16}$ with an iron mass fraction $\rm{\text{Fe-MF}}=0.26^{+0.13}_{-0.12}$ from a detailed \texttt{SuperEarth} structure model. Given the large uncertainties, these values are statistically consistent with zero and should be interpreted as upper limits. The results remains compatible with a wide range of rocky compositions, including Earth-like interiors, and inconsistent with the extreme iron enrichment expected from catastrophic mantle-stripping scenarios. Together with existing thermal phase-curve evidence for a bare-rock surface, our results disfavor giant-impact removal of a primary atmosphere and instead support volatile-poor formation and/or efficient erosion under intense irradiation. Combined with its high emission spectroscopy metric (ESM\(\,=28\)) and the substantial JWST investment already allocated ($50.3$~h across MIRI/LRS and NIRSpec), LHS~3844~b stands out as a benchmark for joint interior-surface studies, and upcoming phase-curve spectroscopy can probe surface mineralogy and search for robust spectral features of exposed rock.

Beyond the planet itself, this work demonstrates the capability of CRIRES$^+$ $K$-band gas-cell observations to deliver precise mass measurements of super-Earths around late M dwarfs. LHS~3844~b is the first rocky exoplanet with its mass and orbit measured using CRIRES$^+$ data, demonstrating that near-infrared RVs provide a practical path to characterize small worlds orbiting cool, faint stars. The infrared brightness of such stars, combined with the large collecting area of the VLT, makes CRIRES$^+$ particularly efficient for monitoring very red targets that are challenging to observe with optical spectrographs. While not designed as an extreme-precision RV instrument, CRIRES$^+$ demonstrates that $\sim 3\,$m\,s$^{-1}$ stability in the $K$ band is achievable \citep{Koehler2025}, opening a complementary regime to optical facilities. This capability is especially relevant for planetary systems orbiting active stars, where observations in the near-infrared can mitigate the impact of stellar activity \citep{Reiners2010}. With ongoing and future surveys such as TESS and PLATO continuing to discover nearby small planets around cool dwarfs, CRIRES$^+$ provides an efficient means to obtain the RV follow-up necessary to expand the sample of well-characterized rocky planets and advance our understanding of their formation and evolution.

After submitting this manuscript, we became aware of an independent analysis of the same ESPRESSO dataset by Hacker et al. to measure the mass of LHS~3844~b.

\textbf{Data availability.} The CRIRES$^+$ and ESPRESSO RV measurements listed in Tables~\ref{table:crires_rvs} and \ref{table:espresso_rvs} are only available in electronic form at the CDS via anonymous ftp to \url{cdsarc.u-strasbg.fr\ (130.79.128.5)} or via \url{http://cdsweb.u-strasbg.fr/cgi-bin/qcat?J/A+A/}.

\begin{acknowledgements}
We thank the anonymous referee for carefully reading our manuscript and for providing constructive comments that have substantially improved its quality.
E.N. acknowledges the support from the Deutsches Zentrum für Luft- und Raumfahrt (DLR, German Aerospace Center) - project number 50OP2502. 
L.B.-Ch., A.D.R., and N.P. acknowledge support by the Knut and Alice Wallenberg Foundation (grant 2018.0192).
A.D.R. acknowledges support from  ANID/Fondo ALMA 2024/N°31240064.
O.K. acknowledges support by the Swedish Research Council (grant agreement no. 2023-03667) and the Swedish National Space Agency.
D.C. is supported by the LMU-Munich Fraunhofer-Schwarzschild Fellowship and by the Deutsche Forschungsgemeinschaft (DFG, German Research Foundation) under Germany's Excellence Strategy -- EXC 2094 -- 390783311.
F.L. is co-funded by the European Union (ERC-CoG, EVAPORATOR, Grant agreement No. 101170037).
M.R. acknowledges support from the DFG Priority Program SPP 1992 “Exploring the Diversity of Extrasolar Planets” (DFG PR 36 24602/41).
This work has made use of data from the European Space Agency (ESA) mission
{\it Gaia} (\url{https://www.cosmos.esa.int/gaia}), processed by the {\it Gaia}
Data Processing and Analysis Consortium (DPAC,
\url{https://www.cosmos.esa.int/web/gaia/dpac/consortium}). Funding for the DPAC
has been provided by national institutions, in particular the institutions
participating in the {\it Gaia} Multilateral Agreement.
This paper makes use of data from the MEarth Project, which is a collaboration 
between Harvard University and the Smithsonian Astrophysical Observatory. 
The MEarth Project acknowledges funding from the David and Lucile Packard Fellowship 
for Science and Engineering, the National Science Foundation under grants AST-0807690, 
AST-1109468, AST-1616624 and AST-1004488 (Alan T. Waterman Award), the National Aeronautics 
and Space Administration under Grant No. 80NSSC18K0476 issued through the XRP Program, 
and the John Templeton Foundation.
This work has made use of \texttt{PyAstronomy} by \citet{pya} (publicly available in 
\url{https://github.com/sczesla/PyAstronomy}).
This research has made use of the SIMBAD database, operated at CDS, Strasbourg, France \citep{Wenger2000}.
This work made use of \texttt{tpfplotter} by J. Lillo-Box (publicly available in 
\url{www.github.com/jlillo/tpfplotter}), which also made use of the python packages 
\texttt{astropy} \citep{Astropy2022}, \texttt{lightkurve} \citep{Lightkurve2018}, \texttt{matplotlib} \citep{Hunter2007}, and \texttt{numpy} \citep{Harris2020}.
\end{acknowledgements}

\bibliographystyle{aa}
\bibliography{bibs}

\begin{appendix}
\onecolumn
\section{Additional figures}

\begin{figure}[ht!]
\centering
\includegraphics[width=0.83\textwidth]{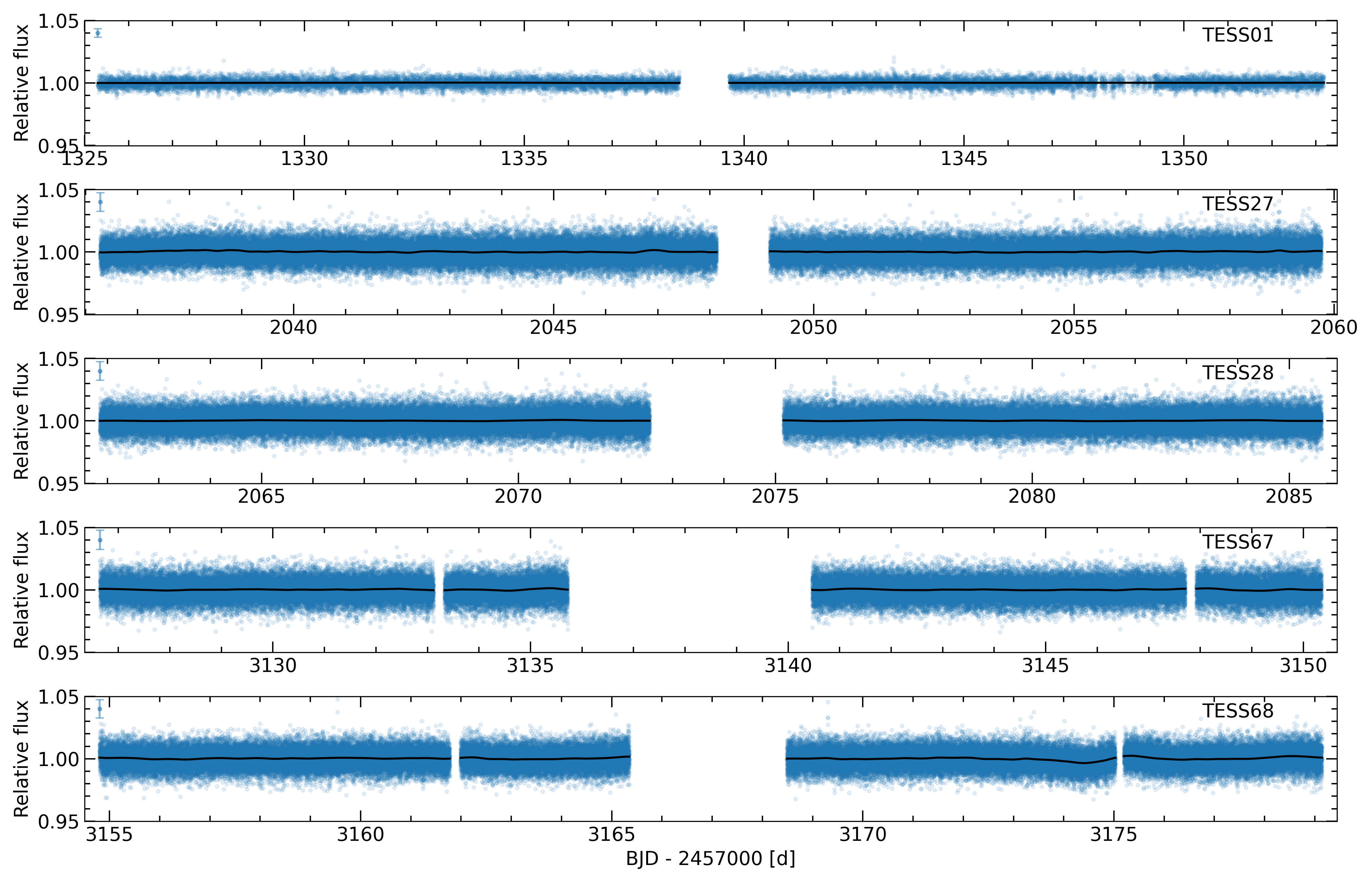}
\caption{\label{figure:juliet_fit_TESS_GPDetrending}
TESS PDCSAP light curves of LHS~3844 for Sectors 1, 27, 28, 67, and 68 (blue dots). 
Overplotted are the best-fit GP models (black lines), which are utilized to detrend the light curves (Sect.~\ref{section:detrending}).}
\end{figure}

\vspace{-20pt}

\begin{figure}[h!]
\centering
\begin{minipage}{0.47\textwidth}
\centering
\includegraphics[width=\linewidth]{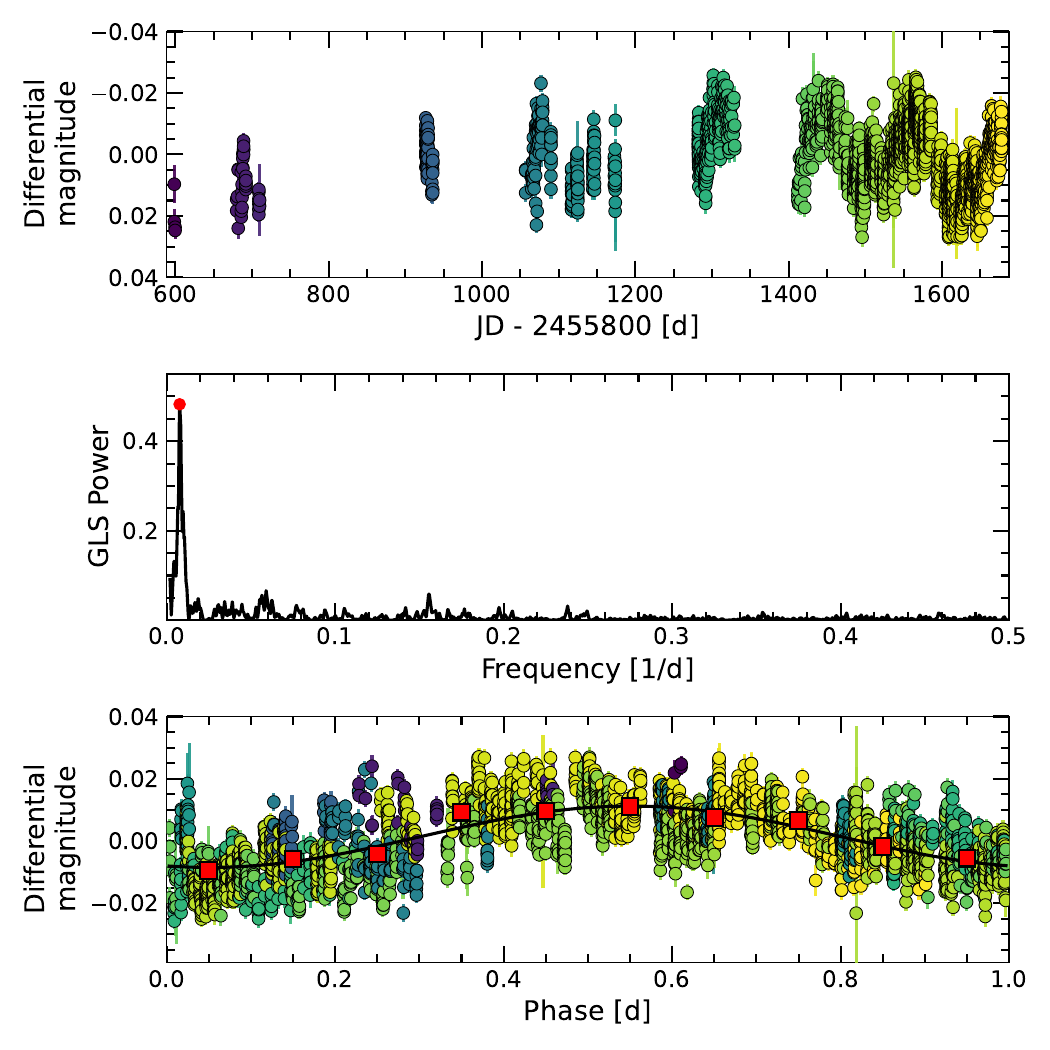}
\caption{Stellar rotation period analysis using MEarth photometric data (Sect.~\ref{sect:stellar_rotation_period}). 
\textit{Top panel:}
MEarth light curve. The color of the data points indicates the observation
epoch. \textit{Middle panel:} GLS periodogram of the MEarth light curve. The red dot indicates the
position of the peak with the highest power at 130.0 days. \textit{Bottom panel:} MEarth light curve 
phase-folded to the period of 130.0 days. The best-fit sinusoidal model is shown in black, while 
the red squares represent the mean magnitude in ten equidistant bins in phase.}
\label{figure:mearth}
\end{minipage}
\hfill
\begin{minipage}{0.47\textwidth}
\centering
\includegraphics[width=\linewidth]{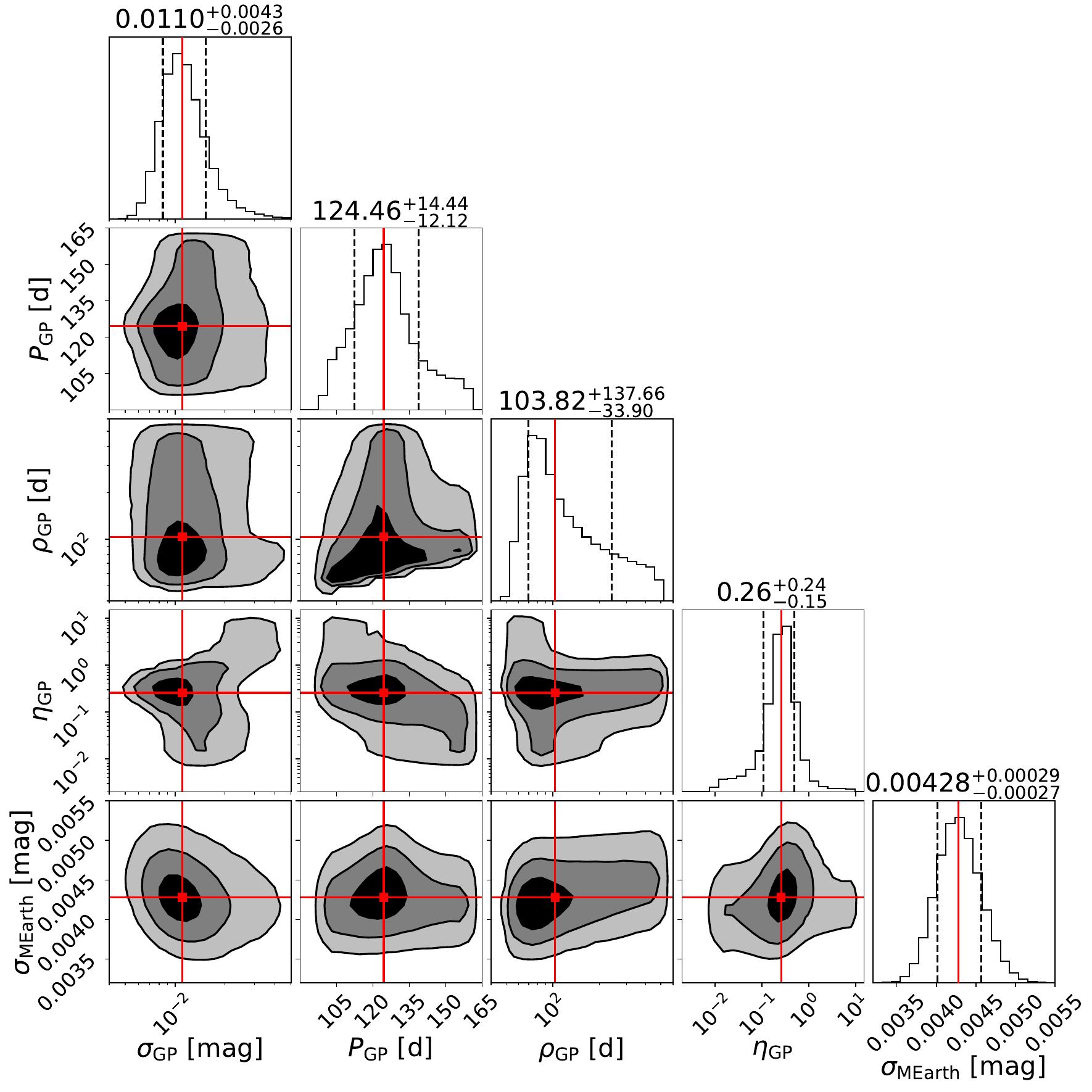}
\caption{Two-dimensional projections of the posterior probability distributions for the GP model fitted to the MEarth photometry (Sect.~\ref{sect:stellar_rotation_period}). The contours represent the 1, 2, and 3$\sigma$ confidence levels.}
\label{figure:spleaf_mearth_corner}
\end{minipage}
\end{figure}

\begin{figure*}
\begin{center}
\includegraphics[width=0.99\textwidth]{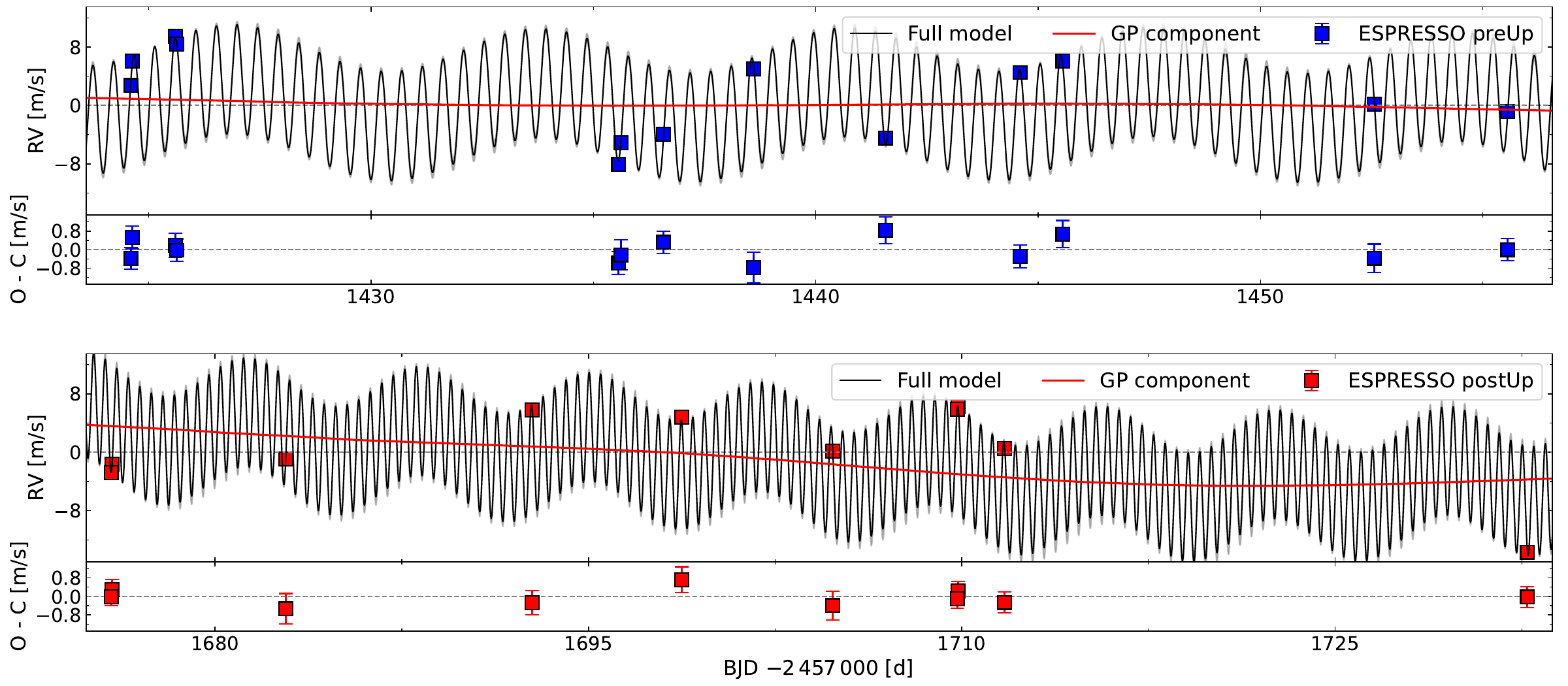}
\caption{\label{figure:espresso_rv_two_planets}
Same as Fig.~\ref{figure:espresso_rv} but for the 2cp + GP model (Sect.~\ref{sect:second_planet}).}
\end{center}
\end{figure*}

\begin{figure}[h!]
\centering
\begin{minipage}{0.48\textwidth}
\centering
\includegraphics[width=\linewidth]{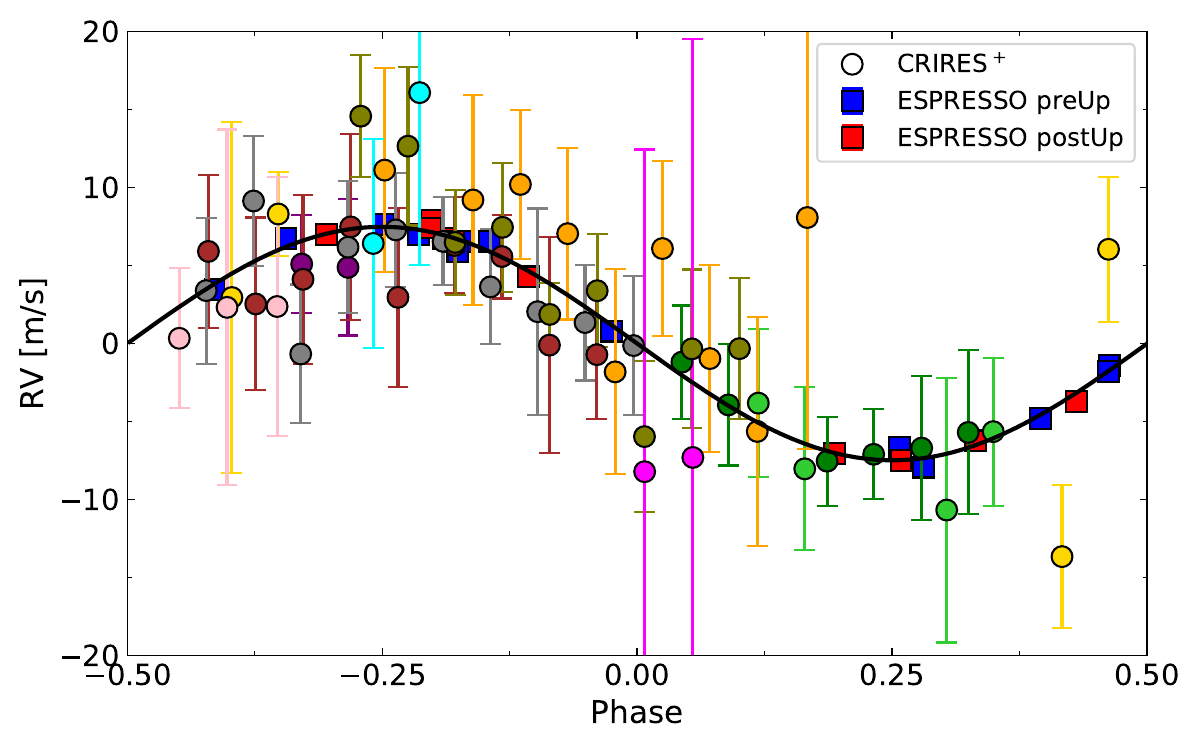}
\caption{Same as Fig.~\ref{figure:rv_phasefold}, but showing the inner planet LHS~3844~b from the 2cp + GP model (Sect.~\ref{sect:second_planet}). The error bars of the ESPRESSO data are smaller than the plotted symbols.}
\label{figure:rv_phasefold_two_planets_1}
\end{minipage}
\hfill
\begin{minipage}{0.50\textwidth}
\centering
\includegraphics[width=\linewidth]{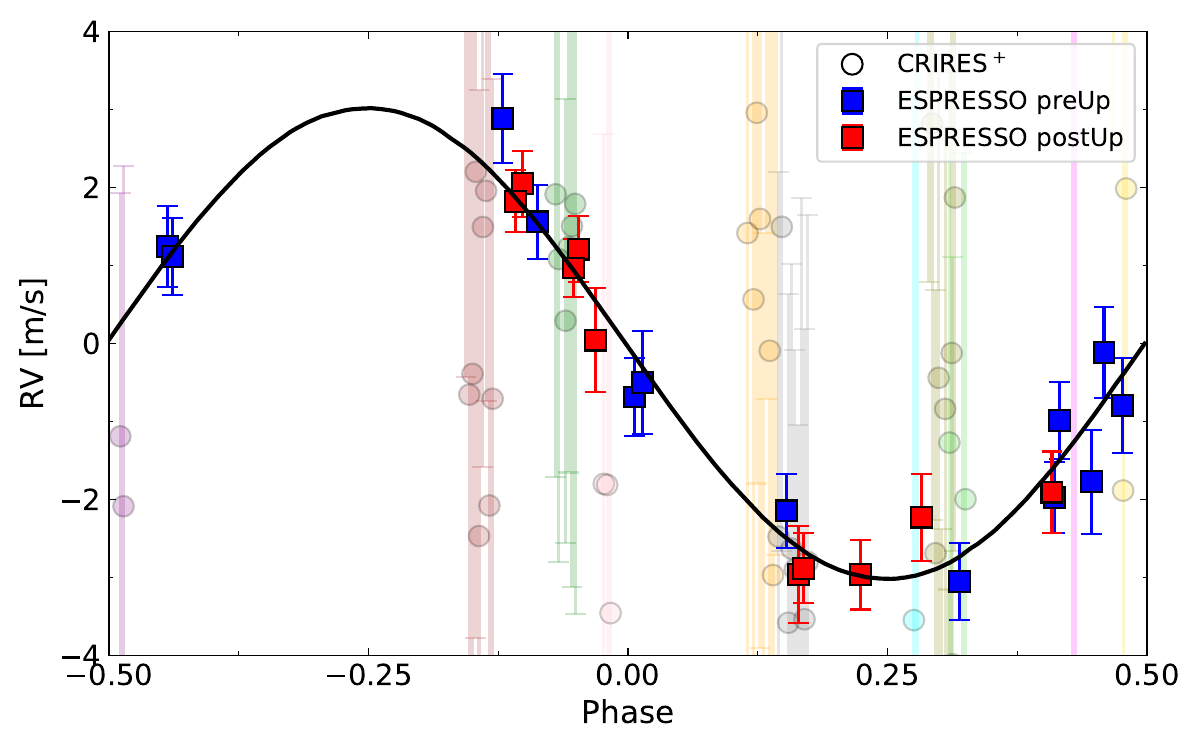}
\caption{Same as Fig.~\ref{figure:rv_phasefold}, but showing the outer planet LHS~3844~c from the 2cp + GP model (Sect.~\ref{sect:second_planet}). For the sake of clarity, the CRIRES$^+$ data are shown with increased transparency.}
\label{figure:rv_phasefold_two_planets_2}
\end{minipage}
\end{figure}

\begin{figure*}
\begin{center}
\includegraphics[width=1.0\textwidth]{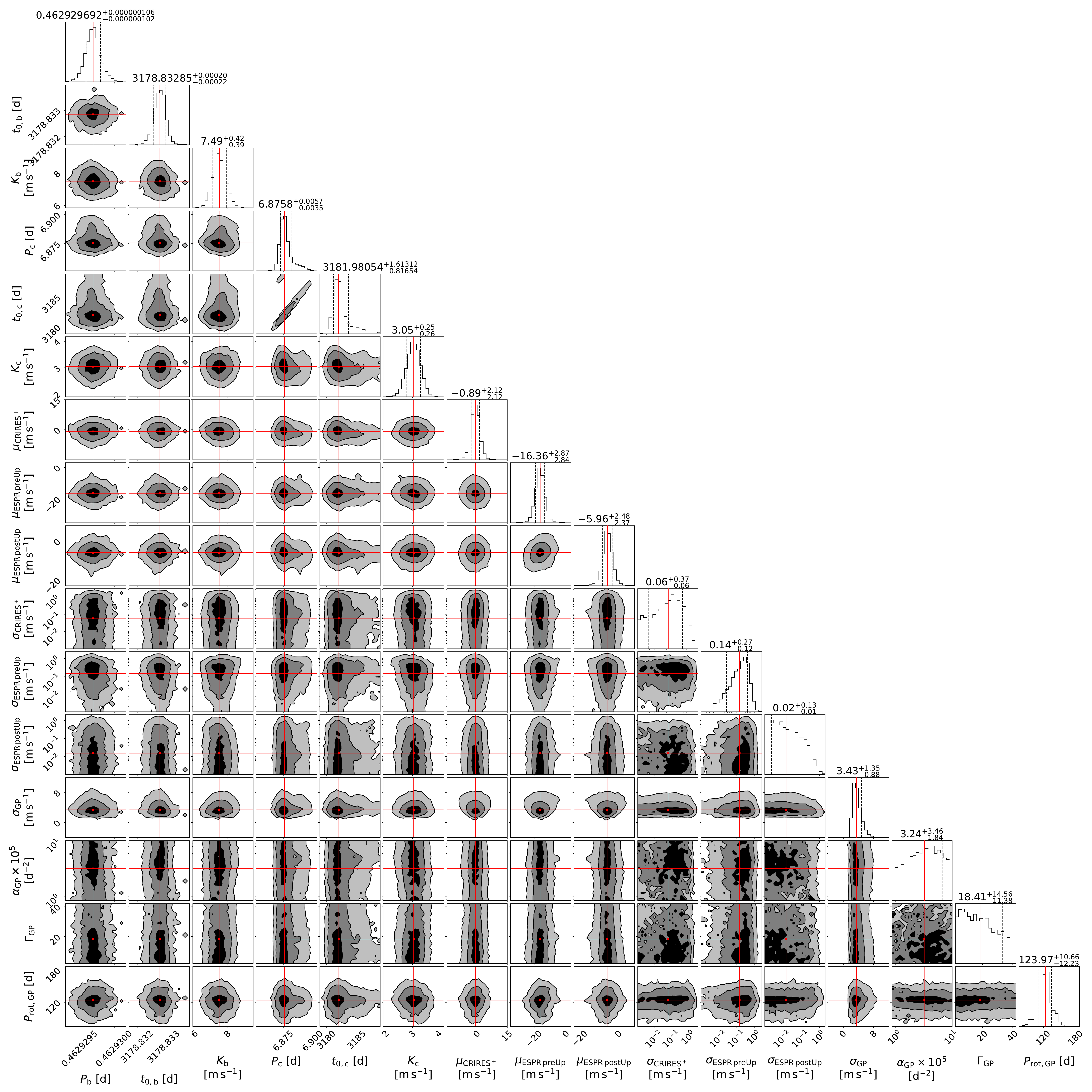}
\caption{\label{figure:corner_rv_only_two_planets}
Two-dimensional projections of the posterior distributions for the 2cp + GP model (Sect.~\ref{sect:second_planet}). The contours represent the 1, 2, and 3$\sigma$ confidence levels.}
\end{center}
\end{figure*}

\begin{figure*}
\begin{center}
\includegraphics[width=1.0\textwidth]{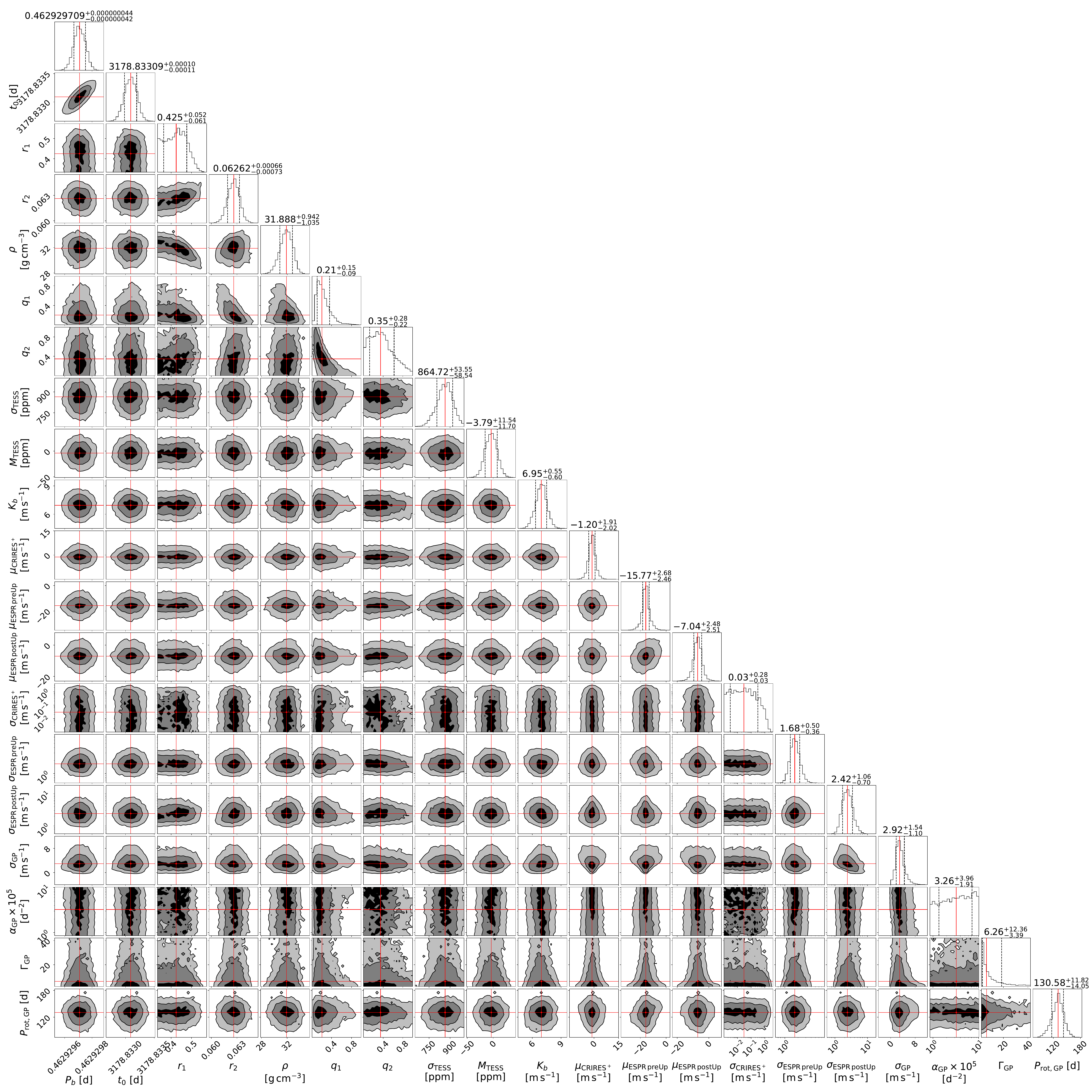}
\caption{\label{figure:tess_phaseFoldet_corner}
Two-dimensional projections of the posterior distributions for the preferred 1cp + GP model of the joint fit (Sect.~\ref{section:joint}). The contours represent the 1, 2, and 3$\sigma$ confidence levels.}
\end{center}
\end{figure*}
\twocolumn

\begin{figure}[h]
\begin{center}
\includegraphics[width=0.49\textwidth]{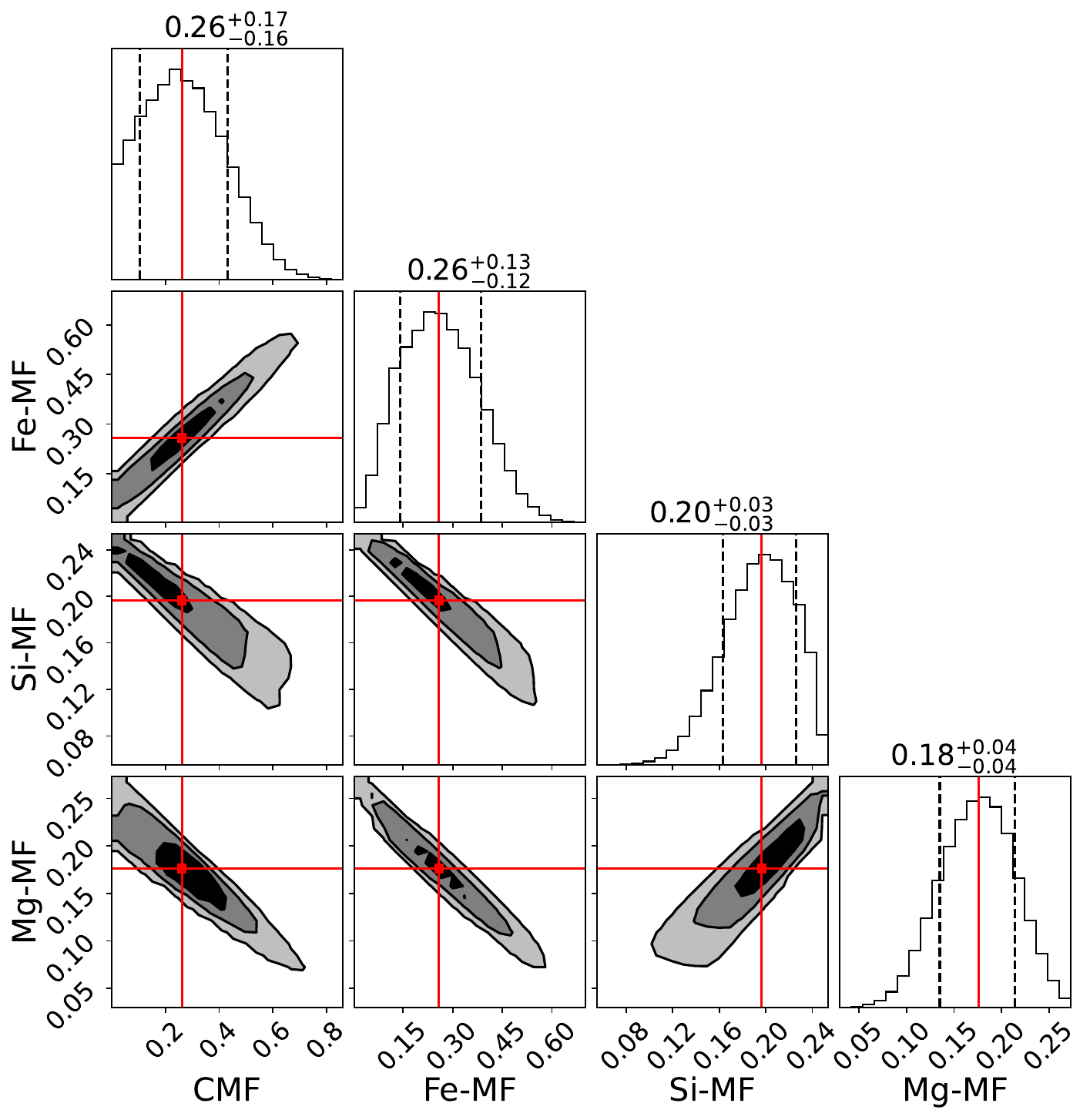}
\caption{\label{figure:exopie}
Two-dimensional projections of the posterior probability distributions of internal structure parameters for LHS~3844~b obtained with \texttt{SuperEarth}. The contours represent the 1, 2, and 3$\sigma$ confidence levels. The analysis constrains the CMS and the mass fraction of iron, silicon, and magnesium using skewed normal priors on the planetary mass and radius and uniform priors on x$_{\rm Fe}$ and x$_{\rm Si}$ (not shown).}
\end{center}
\end{figure}

\FloatBarrier
\onecolumn
\section{Additional tables}

\begin{minipage}[t]{0.48\textwidth}
\centering
\captionof{table}{Barycentric Julian Dates, radial velocities, and formal uncertainties of LHS~3844 measured with CRIRES$^+$.}
\label{table:crires_rvs}
\begin{tabular}{ccc}
\hline\hline
\noalign{\smallskip}
BJD & RV [m\,s$^{-1}$] & $\sigma_{\rm RV}$ [m\,s$^{-1}$]\\
\noalign{\smallskip}
\hline
\noalign{\smallskip}
2459496.529 & $-5.80$ & 4.76 \\
2459496.550 & $-9.98$ & 5.22 \\
2459496.615 & $-12.53$ & 8.49 \\
\vdots & \vdots & \vdots \\
\hline
\end{tabular}
\tablefoot{The full table is available at CDS.}
\end{minipage}
\hfill
\begin{minipage}[t]{0.48\textwidth}
\centering
\captionof{table}{Same as Table~\ref{table:crires_rvs} but for the ESPRESSO data.}
\label{table:espresso_rvs}
\begin{tabular}{ccc}
\hline\hline
\noalign{\smallskip}
BJD & RV [m\,s$^{-1}$] & $\sigma_{\rm RV}$ [m\,s$^{-1}$]\\
\noalign{\smallskip}
\hline
\noalign{\smallskip}
2458424.601 & $-13.49$ & 0.46 \\
2458424.633 & $-10.15$ & 0.50 \\
2458425.603 & $-6.77$ & 0.52 \\
\vdots & \vdots & \vdots \\
\hline
\end{tabular}
\tablefoot{The full table is available at CDS.}
\end{minipage}

\begin{table}[ht!]
\begin{center}
\caption{Priors and posterior estimates from the GP fit to the MEarth photometry of LHS~3844 (Sect.~\ref{sect:stellar_rotation_period}).}
\label{table:posterior_mearth}
\begin{tabular}{lccc} 
\hline\hline 
\noalign{\smallskip}
Parameter & Prior\tablefootmark{(a)} & Posterior value\tablefootmark{(b)} & Description \\
\noalign{\smallskip}
\hline
\noalign{\smallskip}
\multicolumn{4}{c}{\textit{GP hyperparameters}} \\
\noalign{\smallskip}
$\sigma_{\rm GP, MEarth}$ [mag] & $\mathcal{LU}(10^{-5}, 10^{-1})$ & $0.0110^{+0.0043}_{-0.0026}$ & GP Amplitude \\
\noalign{\smallskip}
$P_{\rm GP, MEarth}$ [d] & $\mathcal{U}(100, 160)$ & $124.46^{+14.44}_{-12.12}$ & Characteristic period \\
\noalign{\smallskip}
$\rho_{\rm GP, MEarth}$ [d] & $\mathcal{LU}(5, 500)$ & $103.82^{+137.66}_{-33.90}$ & Coherence timescale \\
\noalign{\smallskip}
$\eta_{\rm GP, MEarth}$ & $\mathcal{LU}(0.01, 10)$ & $0.26^{+0.24}_{-0.15}$ & Periodic damping \\
\noalign{\smallskip}
\hline
\noalign{\smallskip}
\multicolumn{4}{c}{\textit{Photometry parameter}} \\
\noalign{\smallskip}
$\sigma_{\rm MEarth}$ [mag] & $\mathcal{LU}(10^{-7}, 10^{-1})$ & $0.00428^{+0.00029}_{-0.00027}$ & Photometric jitter term \\
\noalign{\smallskip}
\hline
\noalign{\smallskip}
\end{tabular}
\end{center}
\tablefoot{
\tablefoottext{a}{The prior distributions are denoted as follows: $\mathcal{U}$ for uniform, and $\mathcal{LU}$ for log-uniform. Specifically, $\mathcal{U}(a,b)$ and $\mathcal{LU}(a,b)$ correspond to uniform and log-uniform distributions between $a$ and $b$, respectively.}
\tablefoottext{b}{Posterior values are given as medians with 68\,\% credibility intervals.}
}
\end{table}

\begin{table}
\begin{center}
\small
\caption{Priors and posterior estimates from the joint fit of the TESS photometry and RV data for the LHS~3844 system (Sect.~\ref{section:joint}).}
\label{table:posterior}
\begin{tabular}{lccc} 
\hline\hline 
\noalign{\smallskip}
Parameter & Prior\tablefootmark{(a)} & Posterior value\tablefootmark{(b)} & Description \\
\noalign{\smallskip}
\hline
\noalign{\smallskip}
\multicolumn{4}{c}{\textit{Stellar parameters}} \\
$\rho_\star$ [g\,cm$^{-3}$] & $\mathcal{N}(31.730, 1.170)$ & $31.888^{+0.943}_{-1.035}$ & Stellar density \\
\noalign{\smallskip}
\hline
\noalign{\smallskip}
\multicolumn{4}{c}{\textit{Planet parameters LHS~3844~b}} \\
$K_{\rm b}$ [m\,s$^{-1}$] & $\mathcal{U}(0, 10)$ & $6.95^{+0.55}_{-0.60}$ & RV semi-amplitude \\
\noalign{\smallskip}
$P_{\rm b}$ [d] & $\mathcal{U}(0.2315, 0.6945)$ & $0.462929709^{+0.000000044}_{-0.000000042}$ & Orbital period \\
\noalign{\smallskip}
$e_{\rm b}$ & 0 (fixed) & $\dots$  & Orbital eccentricity \\
\noalign{\smallskip}
$\omega_{\rm b}$ [deg] & 90 (fixed) & $\dots$ & Argument of periastron \\
\noalign{\smallskip}
$t_{0, \rm b}$ [BJD - 2\,457\,000] & $\mathcal{U}(3178.5,3179.0)$ & $3178.83309^{+0.00010}_{-0.00011}$ & Mid-transit time \\
\noalign{\smallskip}
$r_{1, \rm b}$ & $\mathcal{U}(0,1)$ & $0.425^{+0.052}_{-0.061}$ & Parameterization related to $p$ and $b$ \\
\noalign{\smallskip}
$r_{2, \rm b}$ & $\mathcal{U}(0,1)$ & $0.06262^{+0.00066}_{-0.00073}$ & Parameterization related to $p$ and $b$ \\
\noalign{\smallskip}
\hline
\noalign{\smallskip}
\multicolumn{4}{c}{\textit{Derived parameters LHS~3844~b}} \\
\noalign{\smallskip}
$a_{\rm b} / R_\star$ & $\dots$ & $7.122^{+0.070}_{-0.078}$ & Scaled semi-major axis \\
\noalign{\smallskip}
$p_{\rm b} = R_{\rm b} / R_\star$ & $\dots$ & $0.06262^{+0.00066}_{-0.00073}$ & Planet-to-star radius ratio \\
\noalign{\smallskip}
$b_{\rm b} = (a_{\rm b} / R_\star) \cos{i_{\rm b}}$ & $\dots$ & $0.137^{+0.078}_{-0.092}$ & Impact parameter \\
\noalign{\smallskip}
$i_{\rm b}$ [deg] & $\dots$ & $88.90^{+0.74}_{-0.65}$ & Orbital inclination \\
\noalign{\smallskip}
$R_{\rm b}$ [$R_\oplus$] & $\dots$ & $1.286^{+0.043}_{-0.044}$ & Planetary radius \\
\noalign{\smallskip}
$m_{\rm b}$ [$M_\oplus$] & $\dots$ & $2.37\pm0.25$ & Planetary mass \\
\noalign{\smallskip}
$\rho_{\rm b}$ [g\,cm$^{-3}$] & $\dots$ & $6.15^{+0.60}_{-0.61}$ & Planet bulk density \\
\noalign{\smallskip}
$a_{\rm b}$ [au] & $\dots$ & $0.00624^{+0.00019}_{-0.00020}$ & Semi-major axis \\
\noalign{\smallskip}
$g_{\rm b}$ [m\,s$^{-2}$] & $\dots$ & $14.10^{+1.19}_{-1.28}$ & Surface gravity \\
\noalign{\smallskip}
$T_{\rm eq, b}$ [K] \tablefootmark{(c)} & $\dots$ & $816\pm14$ & Equilibrium temperature \\
\noalign{\smallskip}
$S_{\rm b}$ [$S_\oplus$] & $\dots$ & $73.67^{+5.14}_{-4.82}$ & Insolation flux \\
\noalign{\smallskip}
$T_{\rm 14,b}$ [min] & $\dots$ & $31.45^{+0.26}_{-0.23}$ & Transit duration \\ 
\noalign{\smallskip}
$T_{\rm 12, b} = T_{\rm 34, b}$ [min] & $\dots$ & $1.90^{+0.06}_{-0.04}$ & Ingress/Egress duration \\
\noalign{\smallskip}
\hline
\noalign{\smallskip}
\multicolumn{4}{c}{\textit{Photometry parameters}} \\
\noalign{\smallskip}
$q_1$ & $\mathcal{U}(0, 1)$ & $0.21^{+0.15}_{-0.09}$ & Limb-darkening parameterization \\
\noalign{\smallskip}
$q_2$ & $\mathcal{U}(0, 1)$ & $0.35^{+0.28}_{-0.22}$ & Limb-darkening parameterization \\
\noalign{\smallskip}
$M_{\rm TESS}$ [ppm] & $\mathcal{N}(0, 0.1)$ & $-3.79^{+11.54}_{-11.70}$ & Relative flux offset\\
\noalign{\smallskip}
$\sigma_{\rm TESS}$ [ppm] & $\mathcal{LU}(10^{-1}, 10^3)$ & $864.72^{+53.55}_{-58.54}$ & Photometric jitter term \\
\noalign{\smallskip}
$D_{\rm TESS}$ & 1 (fixed) & $\dots$ & Dilution factor \\
\noalign{\smallskip}
\hline
\noalign{\smallskip}
\multicolumn{4}{c}{\textit{RV parameters}} \\
\noalign{\smallskip}
$\mu_{\rm CRIRES^+}$ [m\,s$^{-1}$] & $\mathcal{U}(-100, 100)$ & $-1.20^{+1.91}_{-2.03}$ & RV zero-point offset \\
\noalign{\smallskip}
$\sigma_{\rm CRIRES^+}$ [m\,s$^{-1}$] & $\mathcal{LU}(10^{-3}, 10^2)$ & $0.03^{+0.28}_{-0.03}$ & CRIRES$^+$ jitter term \\
\noalign{\smallskip}
$\mu_{\rm ESPRESSO\,preUp}$ [m\,s$^{-1}$] & $\mathcal{U}(-100, 100)$ & $-15.77^{+2.68}_{-2.46}$ & RV zero-point offset \\
\noalign{\smallskip}
$\sigma_{\rm ESPRESSO\,preUp}$ [m\,s$^{-1}$] & $\mathcal{LU}(10^{-3}, 10^2)$ & $1.68^{+0.50}_{-0.36}$ & ESPRESSO preUp jitter term \\
\noalign{\smallskip}
$\mu_{\rm ESPRESSO\,postUp}$ [m\,s$^{-1}$] & $\mathcal{U}(-100, 100)$ & $-7.04^{+2.48}_{-2.51}$ & RV zero-point offset \\
\noalign{\smallskip}
$\sigma_{\rm ESPRESSO\,postUp}$ [m\,s$^{-1}$] & $\mathcal{LU}(10^{-3}, 10^2)$ & $2.42^{+1.06}_{-0.70}$ & ESPRESSO postUp jitter term \\
\noalign{\smallskip}
\hline
\noalign{\smallskip}
\multicolumn{4}{c}{\textit{GP hyperparameters for the RVs}} \\
\noalign{\smallskip}
$\sigma_{\rm GP, RV}$ [m\,s$^{-1}$] & $\mathcal{LU}(10^{-5}, 10^{3})$ & $2.92^{+1.54}_{-1.10}$ & Amplitude of the GP \\
\noalign{\smallskip}
$\alpha_{\rm GP, RV}$ [d$^{-2}$] & $\mathcal{LU}(8.57471 \times 10^{-6},1.02256 \times 10^{-4})$ & $3.25818^{+3.96117}_{-1.90964} \times 10^{-5}$ & Inverse squared-exponential length scale \\
\noalign{\smallskip}
$\Gamma_{\rm GP, RV}$ & $\mathcal{LU}(2.05141,42.1239)$ & $6.26^{+12.36}_{-3.39}$ & Quasi-periodic modulation amplitude \\
\noalign{\smallskip}
$P_{\rm GP, RV}$ [d] & $\mathcal{N}(124.464,14.444)$ & $130.59^{+11.82}_{-14.05}$ & Characteristic period \\
\noalign{\smallskip}
\hline
\noalign{\smallskip}
\end{tabular}
\end{center}
\tablefoot{
\tablefoottext{a}{The prior distributions are denoted as follows: $\mathcal{N}$ for normal, $\mathcal{U}$ for uniform, and $\mathcal{LU}$ for log-uniform. Specifically, $\mathcal{N}(\mu, \sigma^2)$ denotes a normal distribution with mean $\mu$ and variance $\sigma^2$, while $\mathcal{U}(a,b)$ and $\mathcal{LU}(a,b)$ correspond to uniform and log-uniform distributions between $a$ and $b$, respectively.}
\tablefoottext{b}{Posterior values are given as medians with 68\,\% credibility intervals.}
\tablefoottext{c}{The equilibrium temperature assumes a Bond albedo of zero.}
}
\end{table}

\begin{table}
\small
\begin{center}
\caption{Priors and posterior estimates from the RV-only fit for the 2cp + GP model (see Sect.~\ref{sect:second_planet}).}
\label{table:posterior_second_planet}
\begin{tabular}{lccc} 
\hline\hline 
\noalign{\smallskip}
Parameter & Prior\tablefootmark{(a)} & Posterior value\tablefootmark{(b)} & Description \\
\noalign{\smallskip}
\hline
\noalign{\smallskip}
\multicolumn{4}{c}{\textit{Planet parameters LHS~3844~b}} \\
\noalign{\smallskip}
$K_{\rm b}$ [m\,s$^{-1}$] & $\mathcal{U}(0, 10)$ & $7.49^{+0.42}_{-0.39}$ & RV semi-amplitude \\\noalign{\smallskip}
$P_{\rm b}$ [d] & $\mathcal{N}(0.462929709, 1.23 \times 10^{-7})$ & $0.46292969\pm00000010$ & Orbital period \\
\noalign{\smallskip}
$e_{\rm b}$ & 0 (fixed) & $\dots$  & Orbital eccentricity \\
\noalign{\smallskip}
$\omega_{\rm b}$ [deg] & 90 (fixed) & $\dots$ & Argument of periastron \\
\noalign{\smallskip}
$t_{0, \rm b}$ [BJD - 2\,457\,000] & $\mathcal{N}(3178.83309293,0.00031288)$ & $3178.83285^{+0.00020}_{-0.00022}$ & Mid-transit time \\
\noalign{\smallskip}
\hline
\noalign{\smallskip}
\multicolumn{4}{c}{\textit{Planet parameters LHS~3844~c}} \\
\noalign{\smallskip}
$K_{\rm c}$ [m\,s$^{-1}$] & $\mathcal{U}(0, 10)$ & $3.05^{+0.25}_{-0.26}$ & RV semi-amplitude \\\noalign{\smallskip}
$P_{\rm c}$ [d] & $\mathcal{U}(2, 10)$ & $6.8758^{+0.0057}_{-0.0035}$ & Orbital period \\
\noalign{\smallskip}
$e_{\rm c}$ & 0 (fixed) & $\dots$  & Orbital eccentricity \\
\noalign{\smallskip}
$\omega_{\rm c}$ [deg] & 90 (fixed) & $\dots$ & Argument of periastron \\
\noalign{\smallskip}
$t_{0, \rm c}$ [BJD - 2\,457\,000] & $\mathcal{U}(t_{0, \rm b}, t_{0, \rm b} + 10)$ & $3181.98^{+1.61}_{-0.82}$ & Mid-transit time \\
\noalign{\smallskip}
\hline
\noalign{\smallskip}
\multicolumn{4}{c}{\textit{RV parameters}} \\
\noalign{\smallskip}
$\mu_{\rm CRIRES^+}$ [m\,s$^{-1}$] & $\mathcal{U}(-100, 100)$ & $-0.89\pm2.12$ & RV zero-point offset \\
\noalign{\smallskip}
$\sigma_{\rm CRIRES^+}$ [m\,s$^{-1}$] & $\mathcal{LU}(10^{-3}, 10^2)$ & $0.06^{+0.37}_{-0.06}$ & CRIRES$^+$ jitter term \\
\noalign{\smallskip}
$\mu_{\rm ESPRESSO\,preUp}$ [m\,s$^{-1}$] & $\mathcal{U}(-100, 100)$ & $-16.36^{+2.87}_{-2.84}$ & RV zero-point offset \\
\noalign{\smallskip}
$\sigma_{\rm ESPRESSO\,preUp}$ [m\,s$^{-1}$] & $\mathcal{LU}(10^{-3}, 10^2)$ & $0.14^{+0.27}_{-0.12}$ & ESPRESSO preUp jitter term \\
\noalign{\smallskip}
$\mu_{\rm ESPRESSO\,postUp}$ [m\,s$^{-1}$] & $\mathcal{U}(-100, 100)$ & $-5.96^{+2.48}_{-2.37}$ & RV zero-point offset \\
\noalign{\smallskip}
$\sigma_{\rm ESPRESSO\,postUp}$ [m\,s$^{-1}$] & $\mathcal{LU}(10^{-3}, 10^2)$ & $0.02^{+0.13}_{-0.01}$ & ESPRESSO postUp jitter term \\
\noalign{\smallskip}
\hline
\noalign{\smallskip}
\multicolumn{4}{c}{\textit{GP hyperparameters for the RVs}} \\
\noalign{\smallskip}
$\sigma_{\rm GP, RV}$ [m\,s$^{-1}$] & $\mathcal{LU}(10^{-5}, 10^{3})$ & $3.43^{+1.35}_{-0.88}$ & Amplitude of the GP \\
\noalign{\smallskip}
$\alpha_{\rm GP, RV}$ [d$^{-2}$] & $\mathcal{LU}(8.57471 \times 10^{-6},1.02256 \times 10^{-4})$ & $3.23926^{+3.45816}_{-1.83541} \times 10^{-5}$ & Inverse squared-exponential length scale \\
\noalign{\smallskip}
$\Gamma_{\rm GP, RV}$ & $\mathcal{LU}(2.05141,42.1239)$ & $18.41^{+14.57}_{-11.38}$ & Quasi-periodic modulation amplitude \\
\noalign{\smallskip}
$P_{\rm GP, RV}$ [d] & $\mathcal{N}(124.464,14.444)$ & $123.98^{+10.66}_{-12.23}$ & Characteristic period \\
\noalign{\smallskip}
\hline
\noalign{\smallskip}
\end{tabular}
\end{center}
\tablefoot{
\tablefoottext{a}{The prior distributions are denoted as follows: $\mathcal{U}$ for uniform, and $\mathcal{LU}$ for log-uniform. Specifically, $\mathcal{U}(a,b)$ and $\mathcal{LU}(a,b)$ correspond to uniform and log-uniform distributions between $a$ and $b$, respectively.}
\tablefoottext{b}{Posterior values are given as medians with 68\,\% credibility intervals.}
}
\end{table}

\begin{table}
\centering
\caption{Bayesian evidences for RV-only model comparisons performed separately on the CRIRES$^+$ and ESPRESSO datasets, following Table~\ref{tab:bayes_rv-only}.}
\begin{tabular}{ccccc}
\hline\hline
\noalign{\smallskip}
& \multicolumn{2}{c}{CRIRES$^+$} & \multicolumn{2}{c}{ESPRESSO} \\
Model\tablefootmark{(a)} & $\ln{Z} $ & $|\Delta \ln{Z}|$ & $\ln{Z} $ & $|\Delta \ln{Z}|$ \\
\noalign{\smallskip}
\hline
\noalign{\smallskip}
\multicolumn{5}{c}{\textit{Without stellar activity modeling}} \\
\noalign{\smallskip}
0p & $-204.76\pm0.19$ & 0 & $-85.82\pm0.31$ & 0 \\ 
1cp & $-180.79\pm0.18$ & 23.97 & $-69.35\pm0.37$ & 16.47 \\ 
1cp + lt & $-189.23\pm0.32$ & 15.53 & $-72.62\pm0.46$ & 13.19 \\
1ep & $-182.85\pm0.25$ & 21.91 & $-71.53\pm0.40$ & 14.29 \\ 
1ep + lt & $-191.66\pm0.36$ & 13.10 & $-75.00\pm0.50$ & 10.82 \\ 
\noalign{\smallskip}
\multicolumn{5}{c}{\textit{With stellar activity modeling}} \\
\noalign{\smallskip}
0p + GP & $-203.79\pm0.27$ & 0.97 & $-84.85\pm0.31$ & 0.96 \\
1cp + GP & $-180.10\pm0.25$ & 24.66 & $-69.11\pm 0.36$ & 16.71 \\
1cp + lt + GP & $-188.91\pm0.34$ & 15.85 & $-73.37\pm0.50$ & 12.44 \\  
1ep + GP & $-182.35\pm 0.29$ & 22.41 & $-71.18\pm0.40$ & 14.64 \\
1ep + lt + GP & $-191.88\pm0.40$ & 12.88 & $-75.32\pm0.50$ & 10.50 \\
\hline
\end{tabular}
\label{tab:bayes_rv-only_crires}
\tablefoot{
\tablefoottext{a}{0p: No planet; 1cp: Circular single-planet Keplerian model; 1cp + lt: Circular single-planet Keplerian model with linear trend; 1ep: Eccentric single-planet Keplerian model; 1ep + lt: Eccentric single-planet Keplerian model with linear trend; GP: Gaussian process with squared-exponential kernel.}
}
\end{table}

\begin{table}
\begin{center}
\caption{Posterior estimates from RV-only analyses performed independently on the CRIRES$^+$ and ESPRESSO datasets (Sect.~\ref{sect:independent_rv_analyses}). }
\label{table:posterior_only_crires}
\begin{tabular}{lcccc} 
\hline\hline 
\noalign{\smallskip}
Instrument & CRIRES$^+$ & CRIRES$^+$ & ESPRESSO & ESPRESSO \\
\noalign{\smallskip}
Model & 1cp & 1cp + GP & 1cp & 1cp + GP \\
\noalign{\smallskip}
Parameter & Posterior value & Posterior value & Posterior value & Posterior value \\
\noalign{\smallskip}
\hline
\noalign{\smallskip}
\multicolumn{5}{c}{\textit{Planet parameters}} \\
$K_{\rm b}$ [m\,s$^{-1}$] & $6.84^{+0.82}_{-0.83}$ & $7.19^{+1.00}_{-0.94}$ & $6.78^{+0.65}_{-0.64}$ & $6.75^{+0.68}_{-0.65}$ \\
\noalign{\smallskip}
$P_{\rm b}$ [d] & $0.462929707^{+0.000000122}_{-0.000000126}$ & $0.462929709^{+0.000000119}_{-0.000000130}$ & $0.462929704^{+0.000000126}_{-0.000000127}$ & $0.462929716^{+0.000000120}_{-0.000000124}$ \\
\noalign{\smallskip}
$e_{\rm b}$ & 0 (fixed) &  0 (fixed) & 0 (fixed) & 0 (fixed)\\
\noalign{\smallskip}
$\omega_{\rm b}$ [deg] & 90 (fixed) & 90 (fixed) & 90 (fixed) & 90 (fixed)\\
\noalign{\smallskip}
$t_{0, \rm b}$ [BJD - 2\,457\,000] & $3178.83308^{+0.00032}_{-0.00031}$ & $3178.83308^{+0.00030}_{-0.00031}$ & $3178.83309\pm0.00031$ & $3178.83309^{+0.00030}_{-0.00031}$ \\
\noalign{\smallskip}
\hline
\noalign{\smallskip}
\multicolumn{5}{c}{\textit{RV parameters}} \\
\noalign{\smallskip}
$\mu_{\rm CRIRES^+}$ [m\,s$^{-1}$] & $-1.48^{+0.61}_{-0.62}$ & $-1.45^{+1.14}_{-1.03}$ & \dots & \dots \\
\noalign{\smallskip}
$\sigma_{\rm CRIRES^+}$ [m\,s$^{-1}$] & $0.04^{+0.52}_{-0.04}$ & $0.04^{+0.36}_{-0.04}$ & \dots & \dots \\
\noalign{\smallskip}
$\mu_{\rm ESPRESSO\,preUp}$ [m\,s$^{-1}$] & \dots & \dots & $-16.23\pm0.53$ & $-16.16^{+2.94}_{-2.31}$ \\
\noalign{\smallskip}
$\sigma_{\rm ESPRESSO\,preUp}$ [m\,s$^{-1}$] & \dots & \dots & $1.75^{+0.49}_{-0.36}$ & $1.71^{+0.49}_{-0.38}$ \\
\noalign{\smallskip}
$\mu_{\rm ESPRESSO\,postUp}$ [m\,s$^{-1}$] & \dots & \dots & $-6.94^{+1.34}_{-1.31}$ & $-7.08^{+2.49}_{-2.65}$ \\
\noalign{\smallskip}
$\sigma_{\rm ESPRESSO\,postUp}$ [m\,s$^{-1}$] & \dots & \dots & $4.05^{+1.24}_{-0.85}$ & $2.67^{+1.61}_{-0.99}$ \\
\noalign{\smallskip}
\hline
\noalign{\smallskip}
\multicolumn{5}{c}{\textit{GP hyperparameters}} \\
\noalign{\smallskip}
$\sigma_{\rm GP, RV}$ [m\,s$^{-1}$] & \dots & $1.39^{+1.61}_{-1.39}$ & \dots & $2.92^{+3.04}_{-2.89}$ \\
\noalign{\smallskip}
$\alpha_{\rm GP, RV}$ [d$^{-2}$] & \dots & $3.02^{+3.92}_{-1.76} \times 10^{-5}$ & \dots & $3.41^{+3.75}_{-2.00} \times 10^{-5}$ \\
\noalign{\smallskip}
$\Gamma_{\rm GP, RV}$ & \dots & $9.89^{+17.20}_{-6.34}$ & \dots & $6.89^{+14.83}_{-4.00}$ \\
\noalign{\smallskip}
$P_{\rm GP, RV}$ [d] & \dots & $125.83^{+14.84}_{-15.29}$ & \dots & $127.87^{+13.86}_{-16.09}$ \\
\noalign{\smallskip}
\hline
\noalign{\smallskip}
\end{tabular}
\end{center}
\end{table}

\end{appendix}

\end{document}